\documentclass[preprint]{elsarticle}
\pdfoutput=1

\usepackage{graphics}
\usepackage{fullpage}
\usepackage{subfig}
\usepackage{amsmath}	
\usepackage{amssymb}    
\usepackage{bbm}
\usepackage{overpic}

\newcommand{\be}{\begin{eqnarray}}
 \newcommand{\ee}{\end{eqnarray}}

\providecommand{\norm}[1]{\left\lVert#1\right\rVert}

\journal{Journal of Computational Physics} 

\begin{document}

\begin{frontmatter}



\title{A Split-Step Numerical Method for the Time-Dependent Dirac Equation in 3-D Axisymmetric Geometry}

\author[crm]{Fran\c{c}ois Fillion-Gourdeau}
\ead{filliong@CRM.UMontreal.ca}

\author[carl,crm]{Emmanuel Lorin}
\ead{elorin@math.carleton.ca}

\author[sher,crm]{Andr\'{e} D. Bandrauk}
\ead{andre.bandrauk@usherbrooke.ca}


\address[crm]{Centre de Recherches Math\'{e}matiques, Universit\'{e} de Montr\'{e}al, Montr\'{e}al, Canada, H3T~1J4}
\address[carl]{School of Mathematics and Statistics, Carleton University, Ottawa, Canada, K1S 5B6}
\address[sher]{Laboratoire de chimie th\'{e}orique, Facult\'{e} des Sciences, Universit\'{e} de Sherbrooke, Sherbrooke, Canada, J1K 2R1}

\date{\today}

\begin{abstract}
A numerical method is developed to solve the time-dependent Dirac equation in cylindrical coordinates for 3-D axisymmetric systems. The time evolution is treated by a splitting scheme in coordinate space using alternate direction iteration, while the wave function is discretized spatially on a uniform grid. The longitudinal coordinate evolution is performed exactly by the method of characteristics while the radial coordinates evolution uses Poisson's integral solution, which allows to implement the radial symmetry of the wave function. The latter is evaluated on a time staggered mesh by using Hermite polynomial interpolation and by performing the integration analytically. The cylindrical coordinate singularity problem at $r=0$ is circumvented by this method as the integral is well-defined at the origin. The resulting scheme is reminiscent of non-standard finite differences. In the last step of the splitting, the remaining equation has a solution in terms of a time-ordered exponential, which is approximated to a higher order than the time evolution scheme.  We study the time evolution of Gaussian wave packets, and we evaluate the eigenstates of hydrogen-like systems by using a spectral method. We compare the numerical results to analytical solutions to validate the method. In addition, we present  three-dimensional simulations of relativistic laser-matter interactions, using the Dirac equation.
\end{abstract}


\end{frontmatter}

\section{Introduction}
\label{sec:intro}

The Dirac equation is among the most important equations in theoretical physics and chemistry as it gives a quantum relativistic description of fermions such as electrons and quarks. When these particles are moving at very high velocity or when they are bound by very strong classical fields, the non-relativistic modelling based on the Schr\"odinger equation fails and theoretical investigations should be based on the Dirac equation. The extreme conditions where relativistic effects are important can be found in many areas such as relativistic heavy ion collisions, heavy ion spectroscopy, cosmology, astrophysics, and more recently, in laser-matter interaction (for a review, see \cite{Salamin200641} and references therein) and condensed matter physics \cite{Katsnelson2006}. For this reason, the Dirac equation, coupled to an electromagnetic field, has been studied extensively to evaluate many observables such as electron-positron production, molecule spectra, molecular ionization rates, and others. However, solving this equation remains a very challenging task because of its intricate matrix structure, its unbounded spectrum (the Dirac operator has negative energy states which forbids the use of naive minimization numerical methods \cite{QUA:QUA560250112}) and its multiscales (typically, the electromagnetic field is macroscopic, with a time scale of $t_{\rm E\&M} \sim 0.1 - 100 $ fs $ = 0.1 - 100 \times 10^{-15}$ s, while the electron motion, as in the \textit{zitterbewegung} process, has a time scale of $t_{\rm el} \sim 1$ zs $=10^{-21}$ s, \cite{Greiner:1987}).

Existing approaches to tackle these important problems can usually be classified in three categories. The first one is the analytical method which aims at finding closed-form solutions. Although many important problems were treated in this way \cite{Greiner:1987,bagrov1990exact}, it only allows the study of idealized systems. The second approach is the semi-classical approximation which can be used to study more complex configurations than the analytical method (see \cite{PhysRevLett.89.193001} for instance). However, it is only valid for a certain range of wave function parameters, which may not be realized in the physical system under study. The last one is based on full numerical approximations which in principle, can be used to investigate any physical systems. However, even on the numerical side, the solution of the Dirac equation is a challenging problem: it requires a lot of computer resources \cite{doi:10.1080/09500340210140740} and certain numerical schemes are plagued by numerical artifacts such as the fermion doubling problem \cite{doi:10.1080/09500340210140740,PhysRevD.26.468,Muller1998245,Kogut:1974ag}. Therefore, special cares have to be taken to resolve these issues when solving the Dirac equation numerically for physically relevant systems.

Among the most successful numerical methods solving the Dirac equation, many are based on a split-step scheme where the Dirac Hamiltonian is separated in several operators. This has been used in conjunction with spectral schemes in \cite{PhysRevA.59.604,Mocken2008868,Mocken2004558,PhysRevA.53.1605,Bauke2011} for the Dirac equation and in \cite{Huang2005761,Bao2004663} to solve the coupled Maxwell-Dirac system of equations. Very accurate results (with spectral convergence) were obtained with these methods. However, the main drawback is that the computation time scales like $O(N \log N)$ (where $N$ is the number of spatial points in the discretization) because a Fourier transform has to be computed at every timestep. ``Real space'' methods were exploited by many people using finite element schemes \cite{Muller1998245,PhysRevLett.54.669} and finite difference schemes (both explicit \cite{0022-3700-16-11-017} and implicit \cite{PhysRevA.40.5548,PhysRevA.79.043418,PhysRevC.71.024904}). However, some of these ``real space'' methods suffer from the fermion doubling problem \cite{PhysRevD.26.468,RevModPhys.55.775} which induces numerical artifacts and can lead to inaccurate solutions. This occurs because the real-time discretization usually modifies the dispersion relation \cite{PhysRevD.26.468} such that travelling wave packets acquire a wrong group velocity.  Consequently, the phase of a travelling wave packet cannot be reproduced accurately by numerical methods suffering from fermion doubling, even when the order of convergence is increased \cite{trefethen:113}.

Recently, a simple but new numerical method was developed which uses a split-step scheme in ``real space'' and the method of characteristics to evolve the wave function in time, while the space discretization is performed with finite volume elements \cite{Lorin2011190,FillionGourdeau2012}. In this setting, exact solutions in coordinate space can be used in most steps of the splitting (by choosing carefully the time increment $\delta t$ and the element size $a$), resulting in a scheme which is free from the fermion doubling problem and which can be parallelized very efficiently \cite{FillionGourdeau2012}. This makes for a very powerful and robust numerical technique which allows to study physical systems in Cartesian coordinates, in any number of dimensions. However, for 3-D systems, the computational cost is still very important and thus, only short time events can be treated in that case (such as heavy ion collisions which last for approximately $ t_{\rm RHIC}\sim 10^{-23}-10^{-22}$~s). For longer events, such as laser-matter interaction (with $ t_{\rm pulse}\sim 10^{-18}-10^{-13}$~s), only 2-D calculations are possible and therefore, different strategies have to be developed to cope with the high computational requirements. One solution is to reduce the 3-D problem to a 2-D problem by using symmetry arguments. In this work, we adopt this point of view and study systems which are azimuthally symmetric.  For this reason, we extend the split-step scheme to solve the Dirac equation in cylindrical coordinates. 

The rationale to consider this coordinate system is twofold. First, many physical systems of interests have an azimuthal symmetry and thus, can be treated in cylindrical coordinates. Two examples of this are heavy ion collisions at zero impact parameter, and laser-atom interaction in a counterpropagating laser configuration.  Second, it reduces the mathematical description of a 3-D system to an equation in 2-D which of course, reduces the computation time significantly. On the other hand, these coordinates introduce new complications in the numerical calculations because the Dirac operator acquires singular terms in the coordinate transformation (terms of the form $1/r$ where $r$ is the radial distance). This complicates the numerical evaluation of this operator on the boundary close to $r=0$ ($\partial \Omega_{r=0,\theta,z}$, where $\Omega_{r,\theta,z}$ is the domain of the wave function and where $r \in \mathbb{R}^{+},\theta \in [0,2\pi]$ and $z \in \mathbb{R}$). This problem has been studied for other equations and many solutions were developed for the Navier-Stokes (and other fluid-like) equations, such as the use of pole conditions \cite{Huang1993254}, shifted mesh \cite{Mohseni2000787} and series expansion close to the singularity \cite{Constantinescu2002165}. A treatment of the singularity for the Schr\"odinger equation in cylindrical coordinates can be found in \cite{senechal2003proceedings} where it is shown that the accuracy of the numerical solution can be improved by writing the differential operator in ``self-adjoint form''. Finally, the Dirac equation with finite difference scheme is treated in \cite{0022-3700-16-11-017} where a filter is applied at very time-step to get rid of spurious oscillations close to $r=0$. In this work, we use another approach which consists of a splitting method analogous to the one presented in \cite{Lorin2011190,FillionGourdeau2012} where alternate dimension iteration is performed. The splitting operators are chosen such that all the singular terms are included in the radial evolution operator. The resulting equation can then be transformed into a set of four 2-D scalar wave equations, expressed in polar coordinates. An integral representation of the solution of these equations can then be found: it is the well-known Poisson formula. The latter can be evaluated by interpolating the wave function spinor components using Hermite polynomials: using this polynomial form allows us to evaluate the integral explicitly. This however entails that a time staggered mesh has to be used: the grid points are shifted by a half space step at every time step. This method allows to obtain an accurate approximation of the solution at $r=0$ while preserving the overall computational performance of the numerical method. It also circumvent the singularity problem at $r=0$ because Poisson's solution is well-defined at that position. Finally, the resulting numerical scheme is very similar to non-standard finite difference schemes used to solve other equations \cite{mickens2000applications}. 


This article is organized as follows. In Section \ref{sec:num_meth}, the numerical method and the discretization of the Dirac equation is presented. The splitting scheme is described in details along with boundary conditions at $r=0$. Section \ref{sec:num_results} contains several numerical results and benchmark tests. The order of convergence of the method is determined numerically by looking at the time evolution of a Gaussian wave packet. The latter is also compared to an analytical solution to verify the validity of the method. Then, some more interesting physical systems are considered. The first one is a Gaussian wave packet immersed in a counterpropagating superintense laser field. It is shown that positive energy states, which can be interpreted as the creation of electron-positron pairs, appear in the numerical solution. In the last part of this section, bound state problems are considered where the eigensolutions are determined from an adaptation of the Feit-Fleck method \cite{Feit1982412} to the relativistic case. This allows us to simulate the interaction of a single electron bound in atoms or molecules with a strong laser field. We conclude in Section \ref{sec:conclu}.

Note that in all equations, the light velocity $c$ and fermion mass $m$ are kept explicitly, allowing to adapt the method easily to natural or atomic units (a.u.).

\section{Numerical Methods}
\label{sec:num_meth}

The main equation considered in this work is the Dirac equation in cylindrical coordinates which can be obtained from the Dirac equation in Cartesian coordinates. The latter is given by \cite{Itzykson:1980rh}
\begin{eqnarray}
i\partial_{t} \psi_{c}(t,\mathbf{x}_{c})  &=& \biggl\{ \alpha_{x} \biggl[ -ic \partial_{x} - eA_{x}(t,\mathbf{x}_{c}) \biggr] + \alpha_{y} \biggl[ -ic \partial_{y} - eA_{y}(t,\mathbf{x}_{c}) \biggr] \nonumber \\
&&  + \alpha_{z} \biggl[ -ic \partial_{z} - eA_{z}(t,\mathbf{x}_{c}) \biggr]  + \beta m c^{2} + e\mathbb{I}_{4}V(t,\mathbf{x}_{c}) \biggr\} \psi_{c}(t,\mathbf{x}_{c}),
\label{eq:dirac_cartesian}
\end{eqnarray}
where $\psi_{c}(t,\mathbf{x}_{c}) \in L^{2}(\mathbb{R}^{3}) \otimes \mathbb{C}^{4}$ is the time and coordinate ($\mathbf{x}_{c} = (x,y,z)$) dependent four-spinor, $\mathbf{A}(t,\mathbf{x}_{c})$ represents the three space components of the electromagnetic vector potential, $V(t,\mathbf{x}_{c}) = A_{0}(t,\mathbf{x}_{c})$ is the scalar potential, $e$ is the electric charge (obeying $e=-|e|$ for an electron), $\mathbb{I}_{n}$ is the $n$ by $n$ unit matrix and $\alpha_i,\beta$ are the Dirac matrices. This equation describes physically the relativistic dynamics of a single electron subjected to an external electromagnetic field. As usual, the latter is introduced by using the minimal coupling prescription\footnote{The minimal coupling prescription consists of replacing $\partial_{\mu} \rightarrow \partial_{\mu} + ie A_{\mu}$, where $A_{\mu}$ is the electromagnetic potential with Lorentz index $\mu$.}, which allows to preserve the gauge invariance of the equation.  

Throughout this work, the Dirac representation is used where
\begin{eqnarray}
\alpha_{i} := 
\begin{bmatrix}
	0 & \sigma_{i} \\
	\sigma_{i} & 0 
\end{bmatrix}
 \; \; , \; \;
\beta := 
\begin{bmatrix}
	\mathbb{I}_{2} & 0 \\
	0 & -\mathbb{I}_{2} 
\end{bmatrix} .
\label{eq:dirac_mat}
\end{eqnarray}
The $\sigma_{i}$ are the usual $2 \times 2$ Pauli matrices defined as
\begin{eqnarray}
\sigma_{x} := 
\begin{bmatrix}
0 & 1 \\ 1 & 0  
\end{bmatrix}
\;\; \mbox{,} \;\;
\sigma_{y} := 
\begin{bmatrix}
0 & -i \\ i & 0 
\end{bmatrix}
\;\; \mbox{and} \;\;
\sigma_{z} := 
\begin{bmatrix}
1 & 0 \\ 0 & -1 
\end{bmatrix}.
\end{eqnarray}

The Dirac equation in Cartesian coordinates is transformed to cylindrical coordinates by using the following mapping: 
\begin{eqnarray}
\label{eq:cylx}
x&=& r \cos \theta, \\
\label{eq:cyly}
y&=& r \sin \theta, 
\end{eqnarray}
where the radial distance is $r=\sqrt{x^{2}+y^{2}} \in \mathbb{R}^{+}$ and the polar angle is $\theta = \arctan (y/x) \in [0,2\pi]$. Using the chain rule for derivatives, we get\footnote{The vector components of the electromagnetic potential transform as $A_{x} = A_{r}\cos \theta - A_{\theta} \sin \theta$ and $A_{y} = A_{r}\sin \theta + A_{\theta} \cos \theta$ under this coordinate transformation.}:
\begin{eqnarray}
i\partial_{t} \Psi(t,\mathbf{x})  &=& \biggl\{ \tilde{\alpha}_{x} \biggl[ -ic \partial_{r} - eA_{r}(t,\mathbf{x}) \biggr] + \tilde{\alpha}_{y} \biggl[ -ic\frac{1}{r} \partial_{\theta} - eA_{\theta}(t,\mathbf{x}) \biggr]\nonumber \\
&&  + \alpha_{z} \biggl[ -ic \partial_{z} - eA_{z}(t,\mathbf{x}) \biggr]   + \beta m c^{2} + e\mathbb{I}_{4}V(t,\mathbf{x}) \biggr\} \Psi(t,\mathbf{x}),
\label{eq:dirac_cyl}
\end{eqnarray}
where $\mathbf{x} = (r,z,\theta)$ and where the Dirac matrices defined by
\begin{eqnarray}
\tilde{\alpha}_{x} &:=& \alpha _{x} \cos \theta + \alpha_{y} \sin \theta ,\\
\tilde{\alpha}_{y} &:=& -\alpha _{x} \sin \theta + \alpha_{y} \cos \theta ,
\end{eqnarray}
are now space dependent. It is convenient for the following calculation to write these matrices as
\begin{eqnarray}
\tilde{\alpha}_{x} &=& \alpha_{x} e^{\theta \alpha_{x}\alpha_{y}} = \alpha_{x}U(\theta) ,\\
\tilde{\alpha}_{y} &=& \alpha_{y} e^{\theta \alpha_{x}\alpha_{y}} = \alpha_{y}U(\theta),
\end{eqnarray}
where $U(\theta):=e^{\theta \alpha_{x}\alpha_{y}}$. When the system has an azimuthal symmetry, the Dirac equation can be simplified by using separation of variable. This occurs when $A_{\mu}$ has no dependence on the polar angle, i.e. when $A_{\mu}(t,\mathbf{x}) = A_{\mu}(t,r,z)$. In this case, the Hamiltonian commutes with the angular momentum operator. For the remaining of this work, this case will be considered. The $\theta$-dependence can then be factorized by using the following ansatz for the four-spinor with cylindrical symmetry \cite{0305-4470-16-9-024,Kullie2004215}:
\begin{eqnarray}
\label{eq:ansatz}
 \Psi (\mathbf{x},t)  = \left[
\begin{array}{c}
 \psi_{1}(t,r,z) e^{i\mu_{1}\theta} \\
\psi_{2}(t,r,z) e^{i\mu_{2}\theta} \\
\psi_{3}(t,r,z) e^{i\mu_{1}\theta} \\
\psi_{4}(t,r,z) e^{i\mu_{2}\theta}
\end{array} 
\right] = U(-\theta/2)\psi(t,r,z) e^{i j_{z}\theta} 
\end{eqnarray}
where $\mu_{1,2} := j_{z} \mp 1/2$ and where $j_{z}$ is the angular momentum projection on the $z$-axis (it can take one of the values $j_{z} = ...,-\frac{5}{2},-\frac{3}{2},-\frac{1}{2},\frac{1}{2},\frac{3}{2},\frac{5}{2},...$). Inserting this ansatz into Eq. (\ref{eq:dirac_cyl}) and computing the $\theta$-derivative we obtain
\begin{eqnarray}
iU(-\theta/2)\partial_{t}\psi(t,r,z)  &=& \biggl\{ \alpha_{x}U(\theta/2) \biggl[ -ic \partial_{r} -ic\frac{1}{2r} - eA_{r}(t,r,z) \biggr] \nonumber \\
&& + \alpha_{y} U(\theta/2) \biggl[ c\frac{j_{z}}{r} - eA_{\theta}(t,r,z) \biggr] \nonumber \\
&& + \alpha_{z}U(-\theta/2) \biggl[ -ic \partial_{z} - eA_{z}(t,r,z) \biggr]  \nonumber \\
&& + \beta U(-\theta/2) m c^{2} + eU(-\theta/2)V(t,r,z) \biggr\}  \psi(t,r,z).
\label{eq:dirac_cyl2}
\end{eqnarray}
Finally, using the fact that $[U,\alpha_{z}] = [U,\beta] = 0$ and $\alpha_{x,y} U(\theta) = U(-\theta) \alpha_{x,y}$, we can completely factorize the angle dependence and we get
\begin{eqnarray}
i\partial_{t}\psi(t,r,z)  &=& \biggl\{ \alpha_{x} \biggl[ -ic \partial_{r} -ic\frac{1}{2r} - eA_{r}(t,r,z) \biggr] + \alpha_{y}  \biggl[ c\frac{j_{z}}{r} - eA_{\theta}(t,r,z) \biggr] \nonumber \\
&&  + \alpha_{z} \biggl[ -ic \partial_{z} - eA_{z}(t,r,z) \biggr]  + \beta  m c^{2} + eV(t,r,z) \biggr\}  \psi(t,r,z).
\label{eq:dirac_cyl3}
\end{eqnarray}
This equation is the starting point of our analysis and will be solved numerically in subsequent sections. It describes physically the wave function for an electron coupled to an electromagnetic field having an azimuthal symmetry.

\subsection{Boundary conditions at $r=0$}

In this section, boundary conditions at $r=0$ for the wave function are considered ($\left. \psi(t,r,z) \right|_{\partial \Omega_{r=0,\theta,z}}$). When using the cylindrical coordinate system, these boundary conditions are very important to obtain physically relevant solutions. In these coordinates, the Dirac equation may have mathematically allowable solutions singular at the origin and these have to be discarded as being nonphysical\footnote{This is analogous to the ordinary wave equation for which solutions in the radial direction can be given in terms of Bessel and singular Neumann functions. The latter are usually discarded on physical grounds.}.  Moreover, these singular solutions can lead to unstable behavior in numerical calculations, in the neighborhood of $\partial \Omega_{r=0,\theta,z}$ \cite{lewis:2592}.

The boundary conditions can be obtained by considering symmetry and regularity constraints of the wave function. Following \cite{lewis:2592}, we make the following assumption: it is assumed that the wave function $\psi_{c}$ in Cartesian coordinates is infinitely differentiable ($\psi_{c} \in C^{\infty}$) in the whole domain $\mathbb{R}^{3}$, and in particular for $r(x,y)=0$. This assumption, albeit being justified physically, restricts the class of potential that can be considered; for instance, the wave function for a Coulomb potential is singular in $r=0$ and thus, cannot theoretically be considered with our numerical method. Nevertheless, this assumption allows to put constraints on the form of the wave function close to $r=0$, as shown in the following.

\begin{figure}[h]
	\centering
		\includegraphics[width=0.70\textwidth]{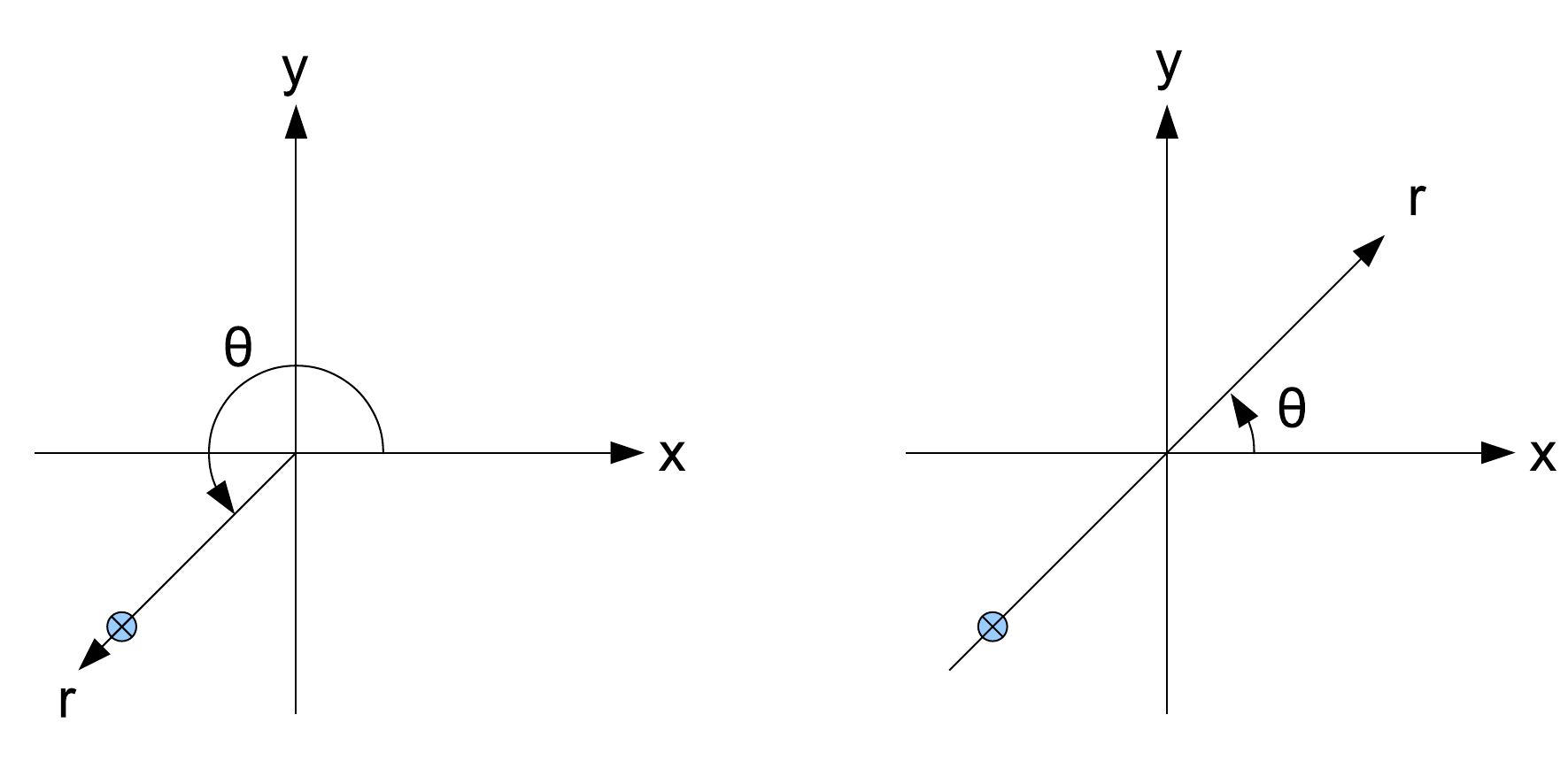}
	\caption{Coordinate transformation symmetry. On the left, the coordinates of an arbitrary point are given in usual polar coordinates while on the right, they are given in the ``transformed'' polar coordinates where $r \in \mathbb{R}$ and $\theta \in [0,\pi]$. }
	\label{fig:coord_sym}
\end{figure}

As demonstrated in \cite{lewis:2592}, the argument starts with the observation that the coordinate transformation in Eqs. (\ref{eq:cylx}) and (\ref{eq:cyly}) is invariant under the symmetry transformation given by $r\rightarrow r'=-r$ and $\theta \rightarrow \theta'= \theta + \pi$. This implies that the wave function in cylindrical coordinates obeys
\begin{eqnarray}
\Psi(t,r,\theta,z) = \Psi(t,-r,\theta + \pi,z). 
\end{eqnarray}
This equivalence is shown in Fig. \ref{fig:coord_sym}.

Using this result in Eq. (\ref{eq:ansatz}), we get the following relations on each spinor components:
\begin{eqnarray}
\label{eq:symm1}
\psi_{1}(t,r,z) &=& (-1)^{|\mu_{1}|} \psi_{1}(t,-r,z) ,  \\
\label{eq:symm2}
\psi_{2}(t,r,z) &=& (-1)^{|\mu_{2}|} \psi_{2}(t,-r,z) ,\\
\label{eq:symm3}
\psi_{3}(t,r,z) &=& (-1)^{|\mu_{1}|} \psi_{3}(t,-r,z) ,\\
\label{eq:symm4}
\psi_{4}(t,r,z) &=& (-1)^{|\mu_{2}|} \psi_{4}(t,-r,z) .
\end{eqnarray}
The values $\mu_{1,2}$ gives the parity of the wave function with respect to $r$ in the ``transformed'' polar coordinates. Thus, when $\mu_{1}$ is even (odd), the components $\psi_{1,3}$ are even (odd) functions of $r$, while when $\mu_{2}$ is even (odd), the components $\psi_{2,4}$ are even (odd) functions of $r$. Note here also that if $\mu_{1}$ is even (odd), then $\mu_{2}$ is odd (even); they are not both even or odd at the same time.

So far, only the symmetry of the coordinate transformation has been used. However, a more explicit form for $\psi$ can be obtained by utilizing the regularity condition. Using the latter, it is demonstrated in \cite{lewis:2592} that the wave function can be written as
\begin{eqnarray}
\label{eq:exp_form1}
\psi_{1}(t,r,z) &=& r^{|\mu_{1}|} f_{1}(t,r^{2},z) ,\\
\label{eq:exp_form2}
\psi_{2}(t,r,z) &=& r^{|\mu_{2}|} f_{2}(t,r^{2},z) ,\\
\label{eq:exp_form3}
\psi_{3}(t,r,z) &=& r^{|\mu_{1}|} f_{3}(t,r^{2},z) ,\\
\label{eq:exp_form4}
\psi_{4}(t,r,z) &=& r^{|\mu_{2}|} f_{4}(t,r^{2},z) .
\end{eqnarray}
where the functions $(f_{i})_{i=1,\cdots,4}$ are regular functions which admit a Taylor expansion in $r^{2}$. These last results allow to understand the behavior of the function $\psi$ as $r \rightarrow 0$. 

The boundary conditions are given for the following two cases.
\begin{enumerate}
 \item When $\mu_{1}$ is even and $\mu_{2}$ is odd:
\begin{eqnarray}
\label{eq:bound_cond1}
\partial_{r} \psi_{1}(t,0,z) &=& 0 ,\\
\label{eq:bound_cond2}
\psi_{2}(t,0,z) &=& 0 ,\\
\label{eq:bound_cond3}
\partial_{r} \psi_{3}(t,0,z) &=& 0 ,\\
\label{eq:bound_cond4}
\psi_{4}(t,0,z) &=& 0 .
\end{eqnarray}
Moreover, when $\mu_{1} \neq 0$, we also have $\psi_{1}(t,0,z) = \psi_{3}(t,0,z) = 0$, which implies that both the function and its derivative are $0$.

 \item When $\mu_{1}$ is odd and $\mu_{2}$ is even:
\begin{eqnarray}
\label{eq:bound_cond5}
\psi_{1}(t,0,z) &=& 0 ,\\
\label{eq:bound_cond6}
\partial_{r}\psi_{2}(t,0,z) &=& 0 ,\\
\label{eq:bound_cond7}
\psi_{3}(t,0,z) &=& 0 ,\\
\label{eq:bound_cond8}
\partial_{r}\psi_{4}(t,0,z) &=& 0 .
\end{eqnarray}
Moreover, when $\mu_{2} \neq 0$, we also have $\psi_{2}(t,0,z) = \psi_{4}(t,0,z) = 0$, which implies that both the function and its derivative are $0$.
\end{enumerate}
These boundary conditions guarantee the continuity of the wave equation. They are Robin conditions. The parity properties and these boundary conditions will be used in the development of the numerical scheme.

\subsection{Operator Splitting and time evolution}

The main goal of this section is to develop a numerical method that allows to evaluate an approximate solution for the Cauchy problem given by a combination of Eq. (\ref{eq:dirac_cyl3}) with the initial condition 
\begin{equation}
 \psi(t_{n},r,z) = \psi^n (r,z), 
\end{equation}
where the wave function is evaluated at time $t_{n}$. As previously discussed in \cite{Lorin2011190,FillionGourdeau2012} for Cartesian coordinates, the wave function can then be evolved to a time $t_{n+1}$ to give $\psi^{n+1}(r,z)$, by using an operator splitting scheme. In this work, a similar approach is adopted for cylindrical coordinates and the following splitting is considered \cite{Lorin2011190,FillionGourdeau2012} (note that we omit the $r,z$ in the wave function argument for notational convenience):
\begin{eqnarray}
\begin{array}{cccll}
\label{eq:split1}
i\partial_{t} \psi^{(1)}(t) &=& \hat{A} \psi^{(1)}(t), & \psi^{(1)}(t_{n}) = \psi^n,  & t \in [t_{n},t_{n+1}) \\
\label{eq:split2}
i\partial_{t} \psi^{(2)}(t) &=&\hat{B} \psi^{(2)}(t), & \psi^{(2)}(t_{n}) = \psi^{(1)}(t_{n+1}), & t \in [t_{n},t_{n+1}) \\
\label{eq:split3}
i\partial_{t} \psi^{(3)}(t) &=& \hat{D} \psi^{(3)}(t), & \psi^{(3)}(t_{n}) = \psi^{(2)}(t_{n+1}), & t \in [t_{n},t_{n+1})  \\
\mbox{and} \;\; \psi^{n+1} &=& \psi^{(3)}(t_{n+1})
\end{array}
\end{eqnarray}
where the upper subscript in parenthesis on the wave function denotes the splitting step number. Note that this splitting scheme leads to an error that scales like $O(\delta t^{2})$, leading to a first order numerical scheme (for more details on the analysis of the method, see \cite{Lorin2011190}). The operators $\hat{A}$, $\hat{B}$ and $\hat{D}$ will be defined subsequently. They are chosen such that an analytical solution can be used to calculate (exactly or approximately) the time evolution for every step of the splitting. 

For each step, the initial condition and time domain are shown explicitly. The method consists of solving each equation independently with an initial condition given by the solution of the previous step. Note also that for every step, the time increment is the same, i.e. $\delta t_n  := t_{n+1} - t_{n}$.  In the following, as the time step is taken constant, we will note $\delta t := \delta t_n$

The operators in the splitting are defined as
\begin{eqnarray}
\label{eq:op_A_s}
\hat{A} &:=&  -ic\alpha_{x} \partial_{r}-ic \alpha_{x} \frac{1}{2r} + c\alpha_{y}\frac{j_{z}}{r}   ,   \\
\label{eq:op_B_s}
\hat{B} &:=&  -i c\alpha_{z} \partial_{z} , \\
\label{eq:op_D_s}
\hat{D} &:=&  \beta m c^{2}+  e\mathbb{I}_{4}V(t,r,z)-e\alpha_{x} A_{r}(t,r,z)  -e  \alpha_{y} A_{\theta}(t,r,z) -e \alpha_{z}A_{z}(t,r,z) .
\end{eqnarray}
This splitting allows to obtain an efficient numerical scheme. For instance, it is shown in the following that the second step can be solved exactly using the method of characteristics, using the technique described in \cite{Lorin2011190,FillionGourdeau2012}. The first step, on the other hand, requires a different technique where interpolation and Poisson's solution are used. The fulfilment of boundary conditions in $r=0$ can be implemented within this framework by using an adapted grid and an appropriate interpolation scheme. In the last step, the solution is given by a ``T-exponential'' which is approximated to second order in $\delta t$. 

\subsubsection{Analytical solution for $\hat{A}$}

The splitting described in the last section is now used to solve the Dirac equation in cylindrical coordinates for systems with azimuthal symmetry. One of the main obstacles to obtain an efficient and accurate numerical scheme is the evaluation of the time evolution of the radial operator $\hat{A}$. The latter has singular coefficients (as $1/r$) coming from the cylindrical coordinate transformation. Therefore, the utilization of a naive finite difference scheme on the mesh described previously is prohibited as it would require the numerical evaluation of $\hat{A}$ at $r=0$, which is obviously undefined. Another approach, suggested in \cite{Mohseni2000787} for fluid-like equation, is to use a shifted mesh where the grid has no points at $r=0$ (the first point is at $r=a/2$). This method however may lead to numerical inaccuracy close to this region in the Dirac equation case. More techniques has been developed to solve similar problems in other equations \cite{Huang1993254,Constantinescu2002165,senechal2003proceedings}. In this work, a new approach is developed based on Poisson's integral solution of the scalar wave equation.

First, the equation obeyed by the wave function is obtained from the definition in Eq. \eqref{eq:op_A_s} and is given by
\begin{eqnarray}
\label{eq:dirac_2d_free}
\begin{cases}
 i\partial_{t} \psi^{(1)}(t,r,z) =  \left[ -ic\alpha_{x} \partial_{r}-ic \alpha_{x} \frac{1}{2r} + c\alpha_{y}\frac{j_{z}}{r}  \right]\psi^{(1)}(t,r,z)\\
 \psi^{(1)}(t_{n},r,z)= \psi^n =: g(r,z), \\
\partial_{t}\psi^{(1)}(t_{n},r,z)= \partial_{t}\psi^{n}=: h(r,z)
\end{cases}
\end{eqnarray}
Writing the last expression explicitly and decoupling the components allows us to write the following Cauchy problem required to evolve the wave function by one timestep, for $t \in [t_{n},t_{n+1})$:
\begin{eqnarray}
\begin{cases}
\label{eq:wav_dir1}
\partial^{2}_{t} \psi^{(1)}_{1}(t,r,z) = c^{2} \left[ \partial^{2}_{r} + \frac{1}{r} \partial_{r} -\frac{\mu_{1}^{2}}{r^{2}}  \right]\psi^{(1)}_{1}(t,r,z) ,\\
\partial^{2}_{t} \psi^{(1)}_{2}(t,r,z) = c^{2} \left[ \partial^{2}_{r} + \frac{1}{r} \partial_{r} -\frac{\mu_{2}^{2}}{r^{2}} \right]\psi^{(1)}_{2}(t,r,z) ,\\
\partial^{2}_{t} \psi^{(1)}_{3}(t,r,z) = c^{2} \left[ \partial^{2}_{r} + \frac{1}{r} \partial_{r} -\frac{\mu_{1}^{2}}{r^{2}} \right]\psi^{(1)}_{3}(t,r,z) ,\\
\label{eq:wav_dir4}
\partial^{2}_{t} \psi^{(1)}_{4}(t,r,z) = c^{2} \left[ \partial^{2}_{r} + \frac{1}{r} \partial_{r} -\frac{\mu_{2}^{2}}{r^{2}} \right]\psi^{(1)}_{4}(t,r,z) ,\\
\psi^{(1)}(t_{n},r,z)=  g(r,z), \\
\partial_{t}\psi^{(1)}(t_{n},r,z)=  h(r,z).
\end{cases}
\end{eqnarray}
The equation obeyed by each wave function component corresponds exactly to the Cauchy problem of the 2-D wave equation in polar coordinates. An analytical solution to this problem is well-known: it is given by  Poisson's formula \cite{polyanin2001handbook,evans2010partial} expressed in polar coordinates (see  \ref{ann:kirchoff} for details on converting Poisson's formula from Cartesian to polar coordinates). Using the fact that the angular dependence of the wave function can be factorized in the form $\Psi(t,r,z,\theta) = e^{i\mu_{1,2} \theta}\psi(t,r,z)$, the solution shown in Eq. \eqref{eq:kirch_cyl} yields, for the spinor components $i=1,\cdots,4$
\begin{eqnarray}
\label{eq:sol_dir_cyl}
 \psi_{i}^{(1)}(t_{n+1},r,z) &=& \frac{1}{2\pi c \delta t}  \int_{B(r,c\delta t)}RdRd\theta \frac{1}{\sqrt{c^{2} \delta t^{2} - [R^{2} + r^{2} - 2Rr \cos(\theta)]}} \nonumber \\
 && \times \biggl\{ \cos(\mu \theta) \biggl[ g_{i}(R,z) + \delta th_{i}(R,z)   + [R-r\cos(\theta)]\partial_{R}g_{i}(R,z) \biggr] \nonumber \\
 && -\sin(\mu \theta) \frac{r}{R}\mu \sin(\theta)g_{i}(R,z) \biggr\} ,
 \end{eqnarray}
where $\mu = \mu_{1}$ for $i=1,3$ and $\mu = \mu_{2}$ for $i=2,4$. Also, $B(r,c \delta t)$ is a disk of radius $c \delta t$ centered at $r$ in the $r-\theta$-plane and corresponds to the integration region (it is depicted in Fig. \ref{fig:int_reg}). As before, $\theta$ is the azimuthal angle. The last equation includes the time derivatives $h_{i}$ which are not convenient for the numerical evaluation. It is possible to discard them by noting that Eq. \eqref{eq:dirac_2d_free}, evaluated at $t=t_{n}$, gives a relation between the different spinor components:
\begin{eqnarray}
h_{1}(r,z) &=& c\left[-\partial_{r} - \frac{\mu_{2}}{r}\right]g_{4}(r,z), \\
h_{2}(r,z) &=& c\left[-\partial_{r} + \frac{\mu_{1}}{r}\right]g_{3}(r,z), \\
h_{3}(r,z) &=& c\left[-\partial_{r} - \frac{\mu_{2}}{r}\right]g_{2}(r,z), \\
h_{4}(r,z) &=& c\left[-\partial_{r} + \frac{\mu_{1}}{r}\right]g_{1}(r,z). 
\end{eqnarray}
Thus, the solution becomes
\begin{eqnarray}
\label{eq:sol_dir_cyl1}
 \psi_{1}^{(1)}(t_{n+1},r,z) &=& \frac{1}{2\pi c \delta t}  \int_{B(r,c\delta t)}RdRd\theta \frac{1}{\sqrt{c^{2} \delta t^{2} - [R^{2} + r^{2} - 2Rr \cos(\theta)]}} \nonumber \\
 && \times \biggl\{ \cos(\mu_{1} \theta) \biggl[ g_{1}(R,z) 
 - c\delta t\left[\partial_{R} + \frac{\mu_{2}}{R}\right]g_{4}(R,z) 
  + [R-r\cos(\theta)]\partial_{R}g_{1}(R,z) \biggr] \nonumber \\
 &&  -\sin(\mu_{1} \theta) \frac{r}{R}\mu_{1} \sin(\theta)g_{1}(R,z) \biggr\} ,\\
 \label{eq:sol_dir_cyl2}
 \psi_{2}^{(1)}(t_{n+1},r,z) &=& \frac{1}{2\pi c \delta t}  \int_{B(r,c\delta t)}RdRd\theta \frac{1}{\sqrt{c^{2} \delta t^{2} - [R^{2} + r^{2} - 2Rr \cos(\theta)]}} \nonumber \\
 && \times \biggl\{ \cos(\mu_{2} \theta) \biggl[ g_{2}(R,z) 
 - c\delta t\left[\partial_{R} - \frac{\mu_{1}}{R}\right]g_{3}(R,z) 
  + [R-r\cos(\theta)]\partial_{R}g_{2}(R,z) \biggr] \nonumber \\
 &&  -\sin(\mu_{2} \theta) \frac{r}{R}\mu_{2} \sin(\theta)g_{2}(R,z) \biggr\} ,\\
 \label{eq:sol_dir_cyl3}
 \psi_{3}^{(1)}(t_{n+1},r,z) &=& \frac{1}{2\pi c \delta t}  \int_{B(r,c\delta t)}RdRd\theta \frac{1}{\sqrt{c^{2} \delta t^{2} - [R^{2} + r^{2} - 2Rr \cos(\theta)]}} \nonumber \\
 && \times \biggl\{ \cos(\mu_{1} \theta) \biggl[ g_{3}(R,z) 
 - c\delta t\left[\partial_{R} + \frac{\mu_{2}}{R}\right]g_{2}(R,z) 
  + [R-r\cos(\theta)]\partial_{R}g_{3}(R,z) \biggr] \nonumber \\
 &&  -\sin(\mu_{1} \theta) \frac{r}{R}\mu_{1} \sin(\theta)g_{3}(R,z) \biggr\} ,\\
 \label{eq:sol_dir_cyl4}
 \psi_{4}^{(1)}(t_{n+1},r,z) &=& \frac{1}{2\pi c \delta t}  \int_{B(r,c\delta t)}RdRd\theta \frac{1}{\sqrt{c^{2} \delta t^{2} - [R^{2} + r^{2} - 2Rr \cos(\theta)]}} \nonumber \\
 && \times \biggl\{ \cos(\mu_{2} \theta) \biggl[ g_{4}(R,z) 
 - c\delta t\left[\partial_{R} - \frac{\mu_{1}}{R}\right]g_{1}(R,z) 
  + [R-r\cos(\theta)]\partial_{R}g_{4}(R,z) \biggr] \nonumber \\
 &&  -\sin(\mu_{2} \theta) \frac{r}{R}\mu_{2} \sin(\theta)g_{4}(R,z) \biggr\} ,
 \end{eqnarray}
which is free from time derivatives. At this point, there is no approximation involved, that is Eqs. \eqref{eq:sol_dir_cyl1} to \eqref{eq:sol_dir_cyl4} are exact solutions of Eq. \eqref{eq:dirac_2d_free}. In Section \ref{sec:spa_dis}, the integral in these equations will be evaluated explicitly and thus, this solution is at the basis of our discretization method. Finally, is should be noted that if the wave function can be Taylor expanded ($\psi \in C^{\infty} \cap  L^{2}(\mathbb{R}^{3}) \otimes \mathbb{C}^{4}$), these integrals are well-defined for $r\in [0,\infty)$ (including the point $r=0$) and thus, there is no coordinate singularity problem in this formulation. This will be shown in explicit calculation in the following.

\subsubsection{Analytical solution for $\hat{B}$}

For the $z$-coordinate component of the Dirac operator, we have to solve Eq. (\ref{eq:split2}) with $\hat{B}$ defined in Eq. (\ref{eq:op_B_s}) on $\mathbb{R}$. To obtain this solution, we starts by diagonalizing the matrix $\alpha_{z}$ using a unitary transformation. The resulting system of equations becomes four transport equations which can be solved by using the method of characteristics. An inverse transformation brings this solution into the original Dirac matrix representation. The details are shown in \ref{app:z_coord} and the final results is given by
\begin{eqnarray}
\label{eq:sol_ex_3}
 \psi^{(2)}(t_{n+1},r,z) &=& \frac{1}{2} \biggl\{ [\mathbb{I}_{4} + \alpha_{z}] \psi^{(1)}(t_{n+1},r,z-c\delta t) + [\mathbb{I}_{4} - \alpha_{z}] \psi^{(1)}(t_{n+1},r,z+c \delta t)  \biggr\} .
\end{eqnarray}
These solutions represent travelling waves moving in opposite directions at velocity $c$. A similar technique was used in \cite{Lorin2011190,FillionGourdeau2012}. 

\subsubsection{Analytical solution for $\hat{D}$}
The last equation of the splitting is, for $t \in [t_{n},t_{n+1})$
\begin{eqnarray}
i\partial_{t} \psi^{(3)}(t,r,z) &=& \bigl[ \beta m c^{2} + e\mathbb{I}_{4}V(t,r,z)\nonumber \\
&& -e\alpha_{x} A_{r}(t,r,z)  -e  \alpha_{y} A_{\theta}(t,r,z) -e \alpha_{z}A_{z}(t,r,z)  \bigr] \psi^{(3)}(t,r,z).
\end{eqnarray}
The solution is simply given by
\begin{eqnarray}
\label{eq:sol_ex_4}
 \psi^{(4)}(t_{n+1},r,z) &=& \hat{T} \exp \biggl\{ -i \int_{t_{n}}^{t_{n+1}} d\tau \bigl[ \beta m c^{2}  -e\alpha_{x} A_{r}(t,r,z)  -e  \alpha_{y} A_{\theta}(t,r,z) -e \alpha_{z}A_{z}(t,r,z)  \bigr]  \biggr\} \nonumber \\
&& \times \exp \left[-ie \int_{t_{n}}^{t_{n+1}} d\tau V(\tau,r,z)  \right] \psi^{(3)}(t_{n+1},r,z).
\end{eqnarray}
where $\hat{T}$, the time-ordering operator, has been introduced. The latter orders the argument of the exponential function according to their time argument: from the smaller time, on the right to the larger time, on the left. This is required because the operator
\begin{eqnarray}
\hat{h}(t) := \beta m c^{2}  -e\alpha_{x} A_{r}(t,r,z)  -e  \alpha_{y} A_{\theta}(t,r,z) -e \alpha_{z}A_{z}(t,r,z)  
\end{eqnarray}
does not commute with itself when it is evaluated at different times due to its Dirac structure (the commutator $[\hat{h}(t),\hat{h}(t')] \neq 0$). In the following, this $T$-ordering will be omitted ($\hat{T} \exp (...) \rightarrow \exp(...)$) as this approximation results in an error of $O(\delta t^{3})$ \cite{PhysRevA.59.604}, which has the same (or better) accuracy than the method considered. It is possible to approximate the time-ordering operator to higher accuracy by using a higher order splitting, as discussed in \cite{Bandrauk2006346,bandrauk:1185,suzuki:400}.

It is now possible to discretize the equation spatially by using the analytical solution obtained for all operators.

\subsection{Spatial Discretization}
\label{sec:spa_dis}

\begin{figure}
\centering
\includegraphics[width=0.7\textwidth]{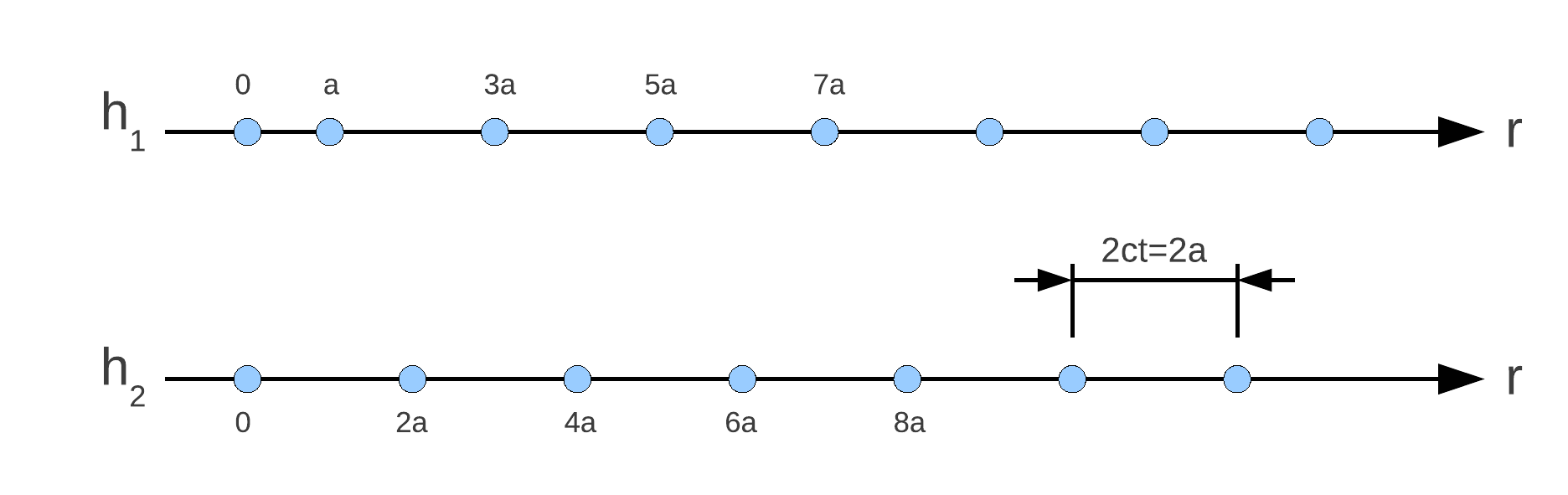}
\caption{Meshes $h_{1}$ and $h_{2}$ used for the discretization in the radial direction. Circles correspond to grid point coordinates $r_{j}^{(h)}$.}
\label{fig:mesh}
\end{figure}

In this numerical method, the space domain is discretized on a grid forming quadrilateral elements with edges of length $\delta r = 2a$ and  $\delta z = a$ (except for elements at $r=0$, as described later), where $\delta r,\delta z$ are the distances between grid points in $r,z$-coordinates and $a$ is the step size. The value of the wave function is evaluated on these mesh points and thus, the discretized wave function and electromagnetic field can be written as
\begin{eqnarray}
\label{eq:discr_psi}
 \psi_{h}(t,j,k) &=& \hat{P}_{h}(j,k)\psi(t,r,z), \\
 &=&  \psi(t,r^{(h)}_{j},z_{k}),\\
\label{eq:discr_A}
 A_{h,r,\theta,z}(t,j,k) &=& \hat{P}_{h}(j,k) A_{r,\theta,z}(t,r,z), \\
 &=& A_{r,\theta,z}(t,r^{(h)}_{j},z_{k}),
\end{eqnarray}
where $\hat{P}_{h}(j,k)$ is an operator that projects a scalar field on point labelled by $(j,k)$ of the grid $h$ ($(j,k) $ is indexing grid points with $j \in \{1,\cdots,N_{r}\}$ and $k \in \{1,\cdots,N_{z}\}$ where $N_{r,z}$ are the number of points in $r,z$ coordinates, respectively) while $ \psi_{h}(t,j,k)$ and $A_{h,r,\theta,z}(t,j,k)$ give the value of the discretized wave-function and electromagnetic field, respectively, on the grid $h$. The explicit definition of coordinates $r^{(h)}_{j}$ and $z_{k}$ depends clearly on the grid used. In this article, two grids will be considered: $h_{1}$ and $h_{2}$. They are depicted in Fig. \ref{fig:mesh} for the radial direction and the point coordinates are defined by 
\begin{eqnarray}
r_{j}^{(h_{1})}&:=&(j-1)2a-a + \delta_{1j}a, \\
r_{j}^{(h_{2})}&:=&(j-1)2a,\\
z_{k} &:=& z_{\rm min} + (k-\frac{1}{2})a,
\end{eqnarray}
with $z_{\rm min}$ the lower domain boundary in $z$-coordinates and $\delta_{1j}$ the Kronecker delta function. These two grids are staggered everywhere except at $r=0$ where both have a grid point. This is very convenient because the exact value of the wave function or its derivative is known at $r=0$, according to the boundary conditions obtained in Eqs. \eqref{eq:bound_cond1} to \eqref{eq:bound_cond8}. Therefore, the boundary conditions at $r=0$ can be implemented exactly on both grids. Another remark is that the grid in the $z$-coordinate is the same for $h_{1}$ and $h_{2}$: only the radial coordinate grid changes. As explained in the next section, these grids are chosen because the solution of the split step in the radial direction requires a time staggered mesh as the mesh changes from $h_{1,2}$ to $h_{2,1}$, respectively, at every time step. We can now discuss the discretization of operators $\hat{A}$, $\hat{B}$ and $\hat{D}$ on these grids.

\subsubsection{Spatial discretization of $\hat{A}$}

The discretization of the operator $\hat{A}$ proceeds by applying the projection operator $\hat{P}_{h}(j,k)$ on Eqs. \eqref{eq:sol_dir_cyl1} to \eqref{eq:sol_dir_cyl4} and by choosing the lattice spacing such that $a = c \delta t$. We obtain
\begin{eqnarray}
\label{eq:proj_sol_dir_cyl1}
 \psi_{1}^{(1)}(t_{n+1},r^{(h)}_{j},z_{k}) &=& \frac{1}{2\pi a}  \int_{B(r^{(h)}_{j},a)}RdRd\theta \frac{1}{\sqrt{a^{2} - [R^{2} + (r^{(h)}_{j})^{2} - 2Rr^{(h)}_{j} \cos(\theta)]}} \nonumber \\
 && \times \biggl\{ \cos(\mu_{1} \theta) \biggl[ g_{1}(R,z_{k}) 
 - a\left[\partial_{R} + \frac{\mu_{2}}{R}\right]g_{4}(R,z_{k}) 
  + [R-r^{(h)}_{j}\cos(\theta)]\partial_{R}g_{1}(R,z_{k}) \biggr] \nonumber \\
 &&  -\sin(\mu_{1} \theta) \frac{r^{(h)}_{j}}{R}\mu_{1} \sin(\theta)g_{1}(R,z_{k}) \biggr\} ,\\
 \label{eq:proj_sol_dir_cyl2}
 \psi_{2}^{(1)}(t_{n+1},r^{(h)}_{j},z_{k}) &=& \frac{1}{2\pi a}  \int_{B(r^{(h)}_{j},a)}RdRd\theta \frac{1}{\sqrt{a^{2}  - [R^{2} + (r^{(h)}_{j})^{2} - 2Rr^{(h)}_{j} \cos(\theta)]}} \nonumber \\
 && \times \biggl\{ \cos(\mu_{2} \theta) \biggl[ g_{2}(R,z_{k}) 
 -a\left[\partial_{R} - \frac{\mu_{1}}{R}\right]g_{3}(R,z_{k}) 
  + [R-r^{(h)}_{j}\cos(\theta)]\partial_{R}g_{2}(R,z_{k}) \biggr] \nonumber \\
 &&  -\sin(\mu_{2} \theta) \frac{r^{(h)}_{j}}{R}\mu_{2} \sin(\theta)g_{2}(R,z_{k}) \biggr\} ,\\
 \label{eq:proj_sol_dir_cyl3}
 \psi_{3}^{(1)}(t_{n+1},r^{(h)}_{j},z_{k}) &=& \frac{1}{2\pi a}  \int_{B(r^{(h)}_{j},a)}RdRd\theta \frac{1}{\sqrt{a^{2}  - [R^{2} + (r^{(h)}_{j})^{2} - 2Rr^{(h)}_{j} \cos(\theta)]}} \nonumber \\
 && \times \biggl\{ \cos(\mu_{1} \theta) \biggl[ g_{3}(R,z_{k}) 
 - a\left[\partial_{R} + \frac{\mu_{2}}{R}\right]g_{2}(R,z_{k}) 
  + [R-r^{(h)}_{j}\cos(\theta)]\partial_{R}g_{3}(R,z_{k}) \biggr] \nonumber \\
 &&  -\sin(\mu_{1} \theta) \frac{r^{(h)}_{j}}{R}\mu_{1} \sin(\theta)g_{3}(R,z_{k}) \biggr\} ,\\
 \label{eq:proj_sol_dir_cyl4}
 \psi_{4}^{(1)}(t_{n+1},r^{(h)}_{j},z_{k}) &=& \frac{1}{2\pi a}  \int_{B(r^{(h)}_{j},a)}RdRd\theta \frac{1}{\sqrt{a^{2}  - [R^{2} + (r^{(h)}_{j})^{2} - 2Rr^{(h)}_{j} \cos(\theta)]}} \nonumber \\
 && \times \biggl\{ \cos(\mu_{2} \theta) \biggl[ g_{4}(R,z_{k}) 
 - a\left[\partial_{R} - \frac{\mu_{1}}{R}\right]g_{1}(R,z_{k}) 
  + [R-r^{(h)}_{j}\cos(\theta)]\partial_{R}g_{4}(R,z_{k}) \biggr] \nonumber \\
 &&  -\sin(\mu_{2} \theta) \frac{r^{(h)}_{j}}{R}\mu_{2} \sin(\theta)g_{4}(R,z_{k}) \biggr\} .
 \end{eqnarray}
These equations give a relation between the wave function at $t=t_{n+1}$ and $t=t_{n}$, assuming that $g$ is defined on all points in the interval $r \in [r^{(h)}_{j}-a,r^{(h)}_{j}+a]$: this is required to perform the integral on the radial coordinate $R$ over the region $B$. However, when the equation is discretized, $g(R,z_{k})$ is not known everywhere on this interval. Rather, we know $\psi_{h'}(t_{n})=g_{h'}$ which is obtained from the preceding time step evolution and which is known only on points of the grid $h'$. It is important to notice that a different grid is used for this initial condition, denoted by $h' \neq h$: this choice will be discussed below. Then, we use the following strategy to perform the radial and angular integrals:
\begin{itemize}
\item The grid points for $h'$ are chosen as $r^{(h')}_{j,j+1}=r^{(h)}_{j} \pm a$. Therefore, they are located on the boundaries of the integration region $B$, on the radial axis. 
\item An approximation of the wave function $g(R,z_{k})$ is obtained by interpolation, such that $g(R,z_{k}) \sim \tilde{g}_{h'}(R,z_{k})$. Here, $\tilde{g}_{h'}$ is an interpolant of $\psi_{h'}(t_{n})$ on the grid $h'$ and thus, can be obtained from $g_{h'}:=\psi_{h'}(t_{n})$. 
\item Then, by replacing $g \rightarrow \tilde{g}$ in Eqs. \eqref{eq:proj_sol_dir_cyl1} to \eqref{eq:proj_sol_dir_cyl4}, it is possible to perform the integrals.
\end{itemize}
By making this choice for $h'$, the two grids at $t_{n+1}$ and $t_{n}$ are staggered and correspond to $h_{1}$ or $h_{2}$ defined previously (if $h=h_{1,2}$, then $h'=h_{2,1}$, respectively). Also, this choice guarantees that only one interpolation polynomial is used throughout the integration region $B$: piecewise polynomial within $B$ is not possible because in this case, the integrals cannot be performed analytically. For this reason, the grid points has to be positioned on the boundary of the integration region, implying that we have to set $c \delta t = a = \delta r/2$. In the next section, it will be shown that this choice is also very convenient for the longitudinal coordinate operator.  

%
The choice of the interpolation method is also important. First, it has to be a polynomial interpolation because in this case, the integral can be evaluated explicitly (this will be shown in the following). In principle, any polynomial interpolation could be used, as long as it obeys the following requirements:
\begin{itemize}
\item It should obey $\tilde{g}_{h} \in C^{1}$ because Eqs. \eqref{eq:sol_dir_cyl1} to \eqref{eq:sol_dir_cyl4} contain a spatial $r$ derivative.
\item It should include information on the derivatives such that boundary conditions at $r=0$ are obeyed exactly by the polynomial.
\item It should be local such that the implementation is relatively easy to perform and the computational performance is preserved.
\item The interpolation error should be bounded by $O(a^{3})$ such that the induced error is smaller or at the same order as the splitting error.
\end{itemize}
In this work, a cubic Hermite interpolation is utilized where the interpolant is given by
\begin{eqnarray}
\tilde{g}_{h'}(r,z_{k})& =&  g_{h'}(j,k)S_{1}(\xi(r)) + g_{h'}(j+1,k)S_{2}(\xi(r)) \nonumber \\
&&+ (r^{(h')}_{j+1}-r^{(h')}_{j})m_{h'}(j,k) S_{3}(\xi(r))+  (r^{(h')}_{j+1}-r^{(h')}_{j})m_{h'}(j+1,k) S_{4}(\xi(r)),
\end{eqnarray}
where $\xi(r) := \frac{r-r^{(h')}_{j}}{r^{(h')}_{j+1}-r^{(h')}_{j}}$ and $m_{h'}$ is the value of the $r$ derivative of $g$ on grid $h'$, that is $m_{h'}(j,k):= \hat{P}_{h'}(j,k)\partial_{r}g(r,z)$. This interpolation uses third order polynomials given by
\begin{eqnarray}
S_{1}(r) &=& 2r^{3} -3r^{2}+1,\\
S_{2}(r) &=& -2r^{3}+3r^{2},\\
S_{3}(r) &=& r^{3}-2r^{2}+r,\\
S_{4}(r) &=& r^{3}-r^{2}. 
\end{eqnarray}
The derivatives are approximated by a symmetric centered difference as 
\begin{eqnarray}
\label{eq:diff1}
m_{h'}(j,k) &\approx & \frac{g_{h'}(j+1,k)-g_{h'}(j-1,k)}{2(r^{(h')}_{j+1}-r^{(h')}_{j-1})}, \\
\label{eq:diff2}
m_{h'}(j+1,k) &\approx & \frac{g_{h'}(j+2,k)-g_{h'}(j,k)}{2(r^{(h')}_{j+2}-r^{(h')}_{j})}.
\end{eqnarray}
It is then a straightforward calculation by using Taylor's approximation to show that $g(r,z_{k}) = \tilde{g}_{h}(r,z_{k}) + O(a^{3})$. By using these last two equations, the interpolant can be written as 
\begin{eqnarray}
\label{eq:interpolant}
\tilde{g}_{h'}(r,z_{k}) =  g_{h'}(j-1,k)P_{1}(r)+g_{h'}(j,k)P_{2}(r) + g_{h'}(j+1,k)P_{3}(r) + g_{h'}(j+2,k)P_{4}(r), 
\end{eqnarray}
where $(P_{n})_{1 \leq n \leq 4}$ are third degree polynomials of the form
\begin{eqnarray}
\label{eq:poly_approx}
P_{n}(r) = \sum_{l=0}^{3}a^{n}_{l}(j) r^{l}.
\end{eqnarray}
The coefficients $a^{n}_{l}(j)$ can be computed explicitly but are not shown here for simplicity. They depend on $r^{(h')}_{j-1}$,$r^{(h')}_{j}$,$r^{(h')}_{j+1}$ and $r^{(h')}_{j+2}$.

\begin{figure}
\centering
\includegraphics[width=0.7\textwidth]{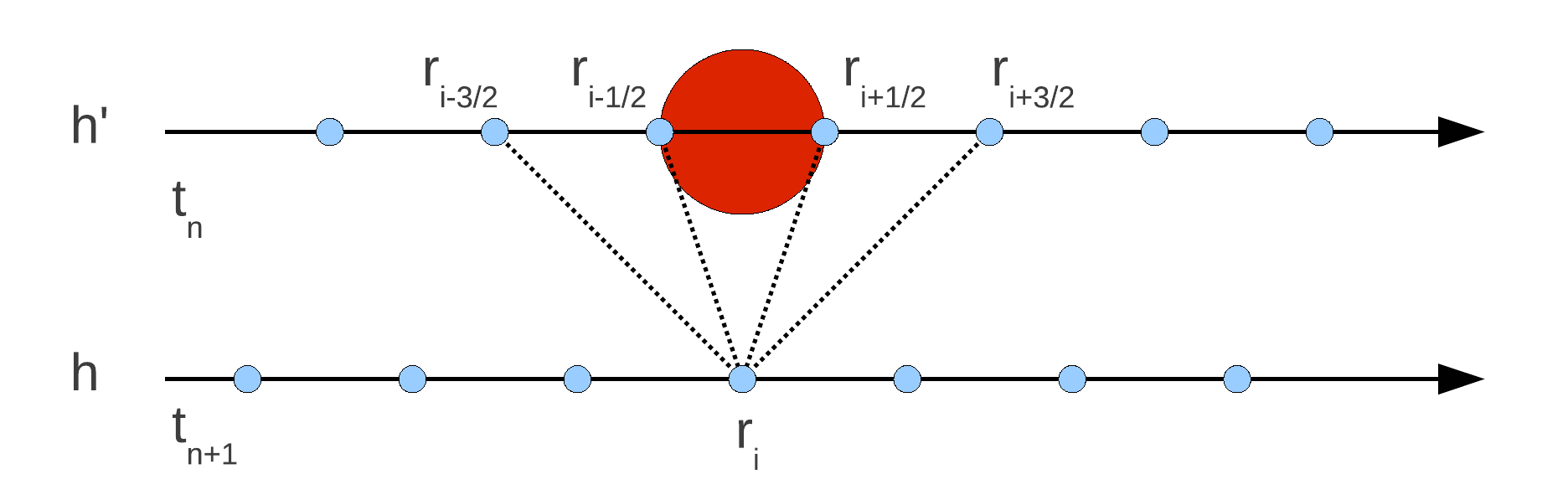}
\caption{Description of the staggered mesh in radial coordinates. The red circle is the integration region $B$. The lines represent the stencil of the scheme.}
\label{fig:stag_mesh}
\end{figure}

\begin{figure}
\centering
\includegraphics[width=0.7\textwidth]{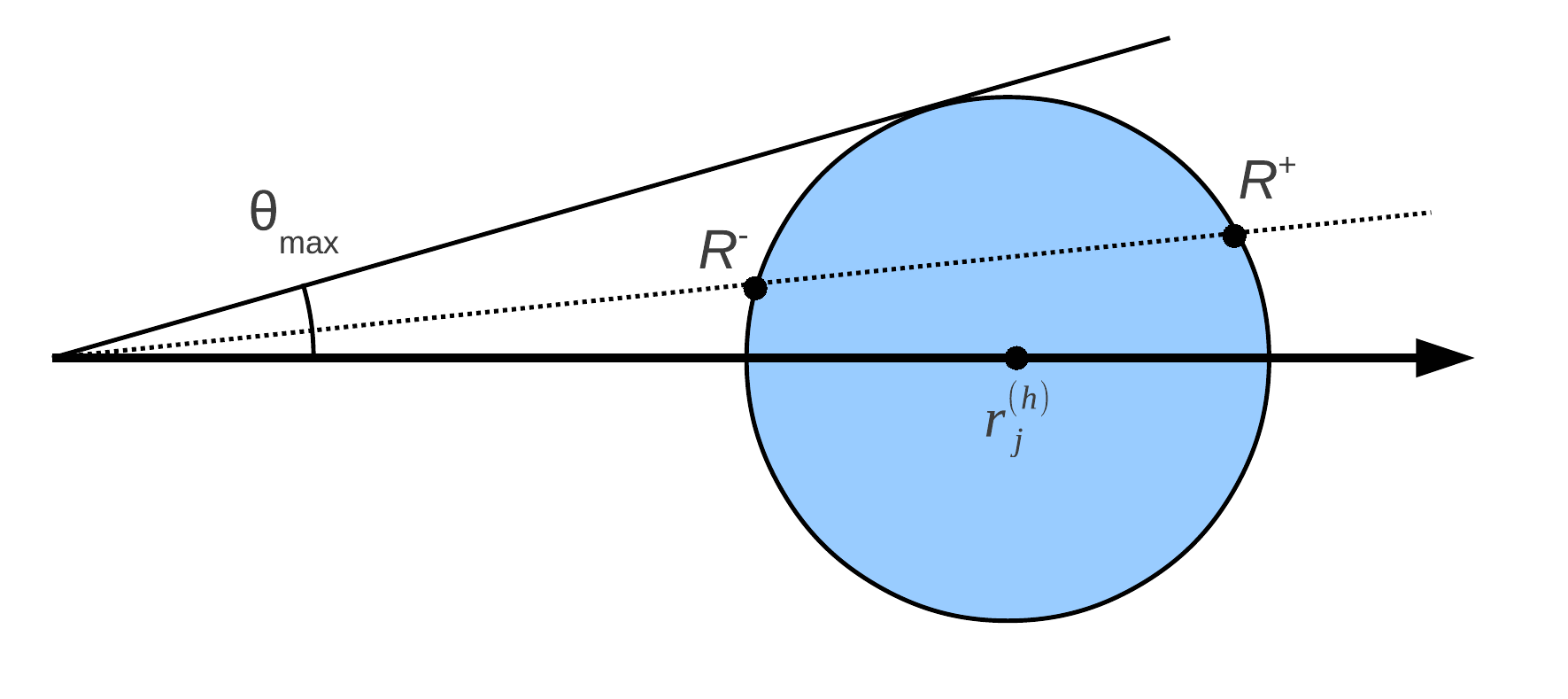}
\caption{Integration region $B$, in blue. The disc is centred on $r=r_{j}^{(h)}$ and has a radius of $a$. The variables $\theta_{\rm max}, R^{+}$ and $R^{-}$ are the boundary of integration.}
\label{fig:int_reg}
\end{figure}

The next step is the substitution of Eq. \eqref{eq:interpolant} into Eqs. \eqref{eq:proj_sol_dir_cyl1} to \eqref{eq:proj_sol_dir_cyl4}, that is we set $g(r,z_{k}) \approx \tilde{g}_{h'}(r,z_{k})$. As a consequence, these equations have integrals of the form 
\begin{eqnarray}
\label{eq:int1}
 I_{l}(r) &=&   \int_{-\theta_{\rm max}}^{\theta_{\rm max}}d\theta \int_{R^{-}}^{R^{+}}dR \frac{f(\theta) R^{l}}{\sqrt{a^{2}  - [R^{2} + r^{2} - 2Rr \cos(\theta)]}} 
 \end{eqnarray}
where $f(\theta)$ is either $\cos(\mu \theta)$, $\cos(\mu \theta)\cos(\theta)$ or $\sin(\mu \theta)\sin(\theta)$ and where the integer $l \in [0,3]$. Also, the integration is performed on the disk $B(r_{j},a)$ centred at $r_{j}$ (the integration region is shown in Fig. \ref{fig:int_reg}) and the integration on the radial coordinate is performed first.  Thus, we have that
\begin{eqnarray}
\theta_{\rm max} &=& \arccos\left( 1- \frac{a^{2}}{2r^{2}}\right), \\
R^{\pm} &=& r\cos(\theta) \pm \sqrt{r^{2} \cos^{2}(\theta) + a^{2} - r^{2}}.
\end{eqnarray} 
We now show that these integrals can be performed analytically. First, we make the change of variable given by
\begin{eqnarray}
y := \frac{R-r\cos(\theta)}{\sqrt{r^{2}\cos^{2}(\theta)+a^{2}-r^{2}}},
\end{eqnarray}  
and the integral becomes
\begin{eqnarray}
\label{eq:int2}
 I_{l}(r) &=&   \int_{-\theta_{\rm max}}^{\theta_{\rm max}}d\theta \int_{-1}^{1}dy \frac{f(\theta) \left[ y\sqrt{r^{2}\cos^{2}(\theta)+a^{2}-r^{2}} + r\cos(\theta)\right]^{l}}{\sqrt{1-y^{2}}} .
 \end{eqnarray}
Now, by using the following result:
\begin{eqnarray}
\int_{-1}^{1} dy \frac{y^{l}}{\sqrt{1-y^{2}}} = \frac{\sqrt{\pi}}{2} \left[1+(-1)^{l} \right] \frac{\Gamma\left(\frac{1}{2} + \frac{l}{2} \right)}{\Gamma\left(1+\frac{l}{2} \right)},
\end{eqnarray}
where $\Gamma$ is the Gamma function, it is possible to conclude that all odd powers of the factor $\sqrt{r^{2}\cos^{2}(\theta)+a^{2}-r^{2}}$ vanish in Eq. \eqref{eq:int2}. As a consequence, after the integration on $R$, Eqs. \eqref{eq:proj_sol_dir_cyl1} to \eqref{eq:proj_sol_dir_cyl4} only contain terms with integer power of $\cos(\theta)$,$\sin(\theta)$,$\cos(\mu \theta)$ and $\sin(\mu\theta)$: these can then be integrated easily on angle $\theta$ by using well-known equation for trigonometric function integrations. The explicit result of this procedure is not shown here for simplicity. It was computed by using the Maple symbolic algebra language and is available on request. 

After performing the integrals in Eqs. \eqref{eq:proj_sol_dir_cyl1} to \eqref{eq:proj_sol_dir_cyl4} by using the interpolant $\tilde{g}_{h'}(r,z_{k})$ and the technique described previously, the result can then be written in a form reminiscent of a non-standard finite difference scheme (this is depicted in Fig. \ref{fig:stag_mesh}):
\begin{eqnarray}
\label{eq:fin_sch1}
\psi^{(1)}_{1}(t_{n+1},r^{(h)}_{j},z_{k})& =& A_{1}(j)g_{1,h}(j-3/2,k) + A_{2}(j)g_{1,h}(j-1/2,k)\nonumber \\
&& + A_{3}(j)g_{1,h}(j+1/2,k)+ A_{4}(j)g_{1,h}(j+3/2,k) \nonumber \\
&& + B_{1}(j)g_{4,h}(j-3/2,k) + B_{2}(j)g_{4,h}(j-1/2,k)\nonumber \\
&& + B_{3}(j)g_{4,h}(j+1/2,k)+ B_{4}(j)g_{4,h}(j+3/2,k)\\
\label{eq:fin_sch2}
\psi^{(1)}_{2}(t_{n+1},r^{(h)}_{j},z_{k})& =& A'_{1}(j)g_{2,h}(j-3/2,k) + A'_{2}(j)g_{2,h}(j-1/2,k) \nonumber \\
&&+ A'_{3}(j)g_{2,h}(j+1/2,k)+ A'_{4}(j)g_{2,h}(j+3/2,k) \nonumber \\
&& + B'_{1}(j)g_{3,h}(j-3/2,k) + B'_{2}(j)g_{3,h}(j-1/2,k) \nonumber \\
&&+ B'_{3}(j)g_{3,h}(j+1/2,k)+ B'_{4}(j)g_{3,h}(j+3/2,k)\\
\label{eq:fin_sch3}
\psi^{(1)}_{3}(t_{n+1},r^{(h)}_{j},z_{k})& =& A_{1}(j)g_{3,h}(j-3/2,k) + A_{2}(j)g_{3,h}(j-1/2,k) \nonumber \\
&&+ A_{3}(j)g_{3,h}(j+1/2,k)+ A_{4}(j)g_{3,h}(j+3/2,k) \nonumber \\
&& + B_{1}(j)g_{2,h}(j-3/2,k) + B_{2}(j)g_{2,h}(j-1/2,k) \nonumber \\
&&+ B_{3}(j)g_{2,h}(j+1/2,k)+ B_{4}(j)g_{2,h}(j+3/2,k)\\
\label{eq:fin_sch4}
\psi^{(1)}_{4}(t_{n+1},r^{(h)}_{j},z_{k})& =& A'_{1}(j)g_{4,h}(j-3/2,k) + A'_{2}(j)g_{4,h}(j-1/2,k) \nonumber \\
&&+ A'_{3}(j)g_{4,h}(j+1/2,k)+ A'_{4}(j)g_{4,h}(j+3/2,k) \nonumber \\
&& + B'_{1}(j)g_{1,h}(j-3/2,k) + B'_{2}(j)g_{1,h}(j-1/2,k)\nonumber \\
&& + B'_{3}(j)g_{1,h}(j+1/2,k)+ B'_{4}(j)g_{1,h}(j+3/2,k)
\end{eqnarray}
where the finite difference coefficients $A_{1,2,3,4}(j)$ depends on the radial position on the mesh. Also, all the quantities are written in terms of the grid $h$ such that
\begin{eqnarray}
g_{h}(j-3/2,k) &=& g_{h'}(j-K_{-2},k) \\
g_{h}(j-1/2,k) &=& g_{h'}(j-K_{-1},k) \\
g_{h}(j+1/2,k) &=& g_{h'}(j+K_{1},k) \\
g_{h}(j+3/2,k) &=& g_{h'}(j+K_{2},k) 
\end{eqnarray}
where the increments are defined as
\begin{eqnarray}
\mbox{if}\;\;h'=h_{1}
\begin{cases}
K_{-2} = 2 \\
K_{-1} = 1 \\
K_{1} = 0 \\ 
K_{2} = 1 
\end{cases}
  \;\; \mbox{or} \;\; \mbox{if}\;\;h'=h_{2}
\begin{cases}
K_{-2} = 1 \\
K_{-1} = 0 \\
K_{1} = 1 \\ 
K_{2} = 2 
\end{cases}
\end{eqnarray}
With these results, it is possible to solve numerically the radial part of the splitting involving operator $\hat{A}$, with an accuracy of $O(a^{3})$: the only source of inaccuracy in this case comes from the interpolation error because the integrals are computed analytically. The order of convergence of this scheme will be demonstrated numerically in Section \ref{sec:num_results}.  Note that it could be improved by using higher order Hermite polynomials (quartic or quintic). Then, the scheme described in Eqs. \eqref{eq:fin_sch1} to \eqref{eq:fin_sch4} would include more terms. 

The scheme described so far is valid for $r>0$. When $r=0$, some of the formula are not valid anymore and some simplifications can be performed. Moreover, the boundary conditions have to be considered. For these reasons, this particular case is treated separately in \ref{app:A_rzero}.

To summarize, the implementation of the numerical scheme consists of the following steps:
\begin{itemize}
 \item Evaluate the analytical formula for the coefficients $A,A',B,B'$. This can be performed by using a symbolic mathematical software.
 \item Evaluate numerically the coefficients $A,A',B,B'$ on the grid $h_{1}$ and $h_{2}$. To increase performance, these coefficients are stored in a pre-computed table.
 \item Evolve the wave function in the splitting scheme by using Eqs. \eqref{eq:fin_sch1} to \eqref{eq:fin_sch4}. 
 \end{itemize}

\subsubsection{Spatial discretization of $\hat{B}$}

The spatial discretization of operator $\hat{B}$ proceeds in the same way as in \cite{Lorin2011190,FillionGourdeau2012}. It is possible to obtain the numerical scheme for this operator by applying the projection operator defined in Eqs. (\ref{eq:discr_psi}) and (\ref{eq:discr_A}) to the analytical solution in  Eq. \eqref{eq:sol_ex_3}. It was shown in \cite{Lorin2011190} that a stable scheme can be obtained by choosing a specific relationship between the time and space increments; it should be given by $c \delta t = K a$ for any $K \in \mathbb{N}^{*}$. When this relation is implemented, the numerical solution for this step of the splitting reproduces the analytical solution exactly, up to errors related to the projection on the grid and to the splitting, and thus, the numerical diffusion is minimized. For $K=1$, this is consistent with the previous discussion on the radial operator $\hat{A}$. 

With this choice of space discretization, the wave function becomes $\psi^{(1)}(t_{n+1},r,z \pm c\delta t) \rightarrow \psi_{h}^{(1)}(t_{n+1},j,k+1)$ and Eq. (\ref{eq:sol_ex_3}) becomes:
\begin{eqnarray}
\label{eq:sol_dis_3}
 \psi_{h}^{(2)}(t_{n+1},j,k) &=& \frac{1}{2} \biggl\{ [\mathbb{I}_{4} + \alpha_{z}] \psi_{h}^{(1)}(t_{n+1},j,k-1)  + [\mathbb{I}_{4} - \alpha_{z}] \psi_{h}^{(1)}(t_{n+1},j,k+1)  \biggr\} .
\end{eqnarray}
Again, we stress that this equation solves Eq. (\ref{eq:sol_ex_3}) exactly, even after discretization.

\subsubsection{Spatial discretization of $\hat{D}$}

For the last step involving the operator $\hat{D}$, the discretization proceed in the same way as the other operator: the projection operator is applied on Eq. \eqref{eq:sol_ex_4}. This yields
\begin{eqnarray}
\label{eq:sol_dis_4}
 \psi^{n+1}_{h}(j,k) & =&  \exp \biggl[ -i \beta mc^{2}\delta t -ie \tilde{V}^{n}_{h}(j,k) \nonumber \\
&& \quad \quad +ie\alpha_{x} \tilde{A}^{n}_{h,r}(j,k)  +ie  \alpha_{y} \tilde{A}^{n}_{h,\theta}(j,k) +ie \alpha_{z}\tilde{A}^{n}_{h,z}(j,k)  \biggr] \psi^{(2)}_{h}(t_{n+1},j,k), \\
\label{eq:sol_dis_5}
  & =&  U(j,k)\exp \biggl[ -ie \tilde{V}^{n}_{h}(j,k) \biggr] \psi^{(2)}_{h}(t_{n+1},j,k),
\end{eqnarray}
where we defined
\begin{eqnarray}
\label{eq:field_dis1}
\tilde{A}^{n}_{r,\theta,z}(r,z) &:=& \int_{t_{n}}^{t_{n+1}} d \tau A_{r,\theta,z}(\tau, r,z) ,\\
\label{eq:field_dis2}
\tilde{V}^{n}(r,z) &:=& \int_{t_{n}}^{t_{n+1}} d\tau V(\tau,r,z) .
\end{eqnarray}
The discretization of $\tilde{A}^{n}_{r,\theta,z}$ and $\tilde{V}^{n}$ proceeds as in Eq. (\ref{eq:discr_psi}) using the projection operator. 

The exponential $U(j,k)$ in Eq. (\ref{eq:sol_dis_4}) actually represents a $4\times 4$ matrix. It can be evaluated explicitly by using a well-known result for the exponential of Dirac matrices, that is\footnote{This relation can be derived by using the Taylor expansion of the exponential function and the following Dirac matrices properties for any positive integers $n$: $\beta^{2n} = 1, \; \alpha_{i}^{2n} = 1, \; \beta^{2n+1} = \beta, \; \alpha_{i}^{2n+1} = \alpha_{i}$.}
\begin{eqnarray}
 e^{i(\beta F + \boldsymbol{\alpha} \cdot \mathbf{F} )} = \cos(|F|) + i \frac{(\beta F + \boldsymbol{\alpha} \cdot \mathbf{F} )}{|F|} \sin(|F|)
\end{eqnarray}
where $\mathbf{F} = (F_{1},F_{2},F_{3})$, $F,F_{i}$ are arbitrary scalar functions and $|F| := \sqrt{F^{2} + \mathbf{F} \cdot \mathbf{F}}$. Using the last equation, the exponential can be given by (here we omit the argument $(j,k)$ in the fields)
\begin{eqnarray}
\label{eq:Umat}
 U(j,k) := 
\begin{bmatrix}
 \mathrm{c}(A) - i\frac{mc^{2} \delta t}{A} \mathrm{s}(A) & 0 & i \frac{\tilde{A}_{h,z}}{A} \mathrm{s}(A) & \frac{ \tilde{A}_{h,\theta} + i\tilde{A}_{h,r}}{A}\mathrm{s}(A)\\
0 &  \mathrm{c}(A) - i\frac{mc^{2} \delta t}{A} \mathrm{s}(A) &\frac{ -\tilde{A}_{h,\theta} + i\tilde{A}_{h,r} }{A}\mathrm{s}(A)  & -i \frac{\tilde{A}_{h,z}(\mathbf{i})}{A} \mathrm{s}(A) \\
i \frac{\tilde{A}_{h,z}(\mathbf{i})}{A} \mathrm{s}(A) & \frac{ \tilde{A}_{h,\theta} + i\tilde{A}_{h,r}}{A}\mathrm{s}(A) &  \mathrm{c}(A) + i\frac{mc^{2} \delta t}{A} \mathrm{s}(A) & 0 \\
\frac{ -\tilde{A}_{h,\theta} + i\tilde{A}_{h,r} }{A}\mathrm{s}(A) & -i \frac{\tilde{A}_{h,z}}{A} \mathrm{s}(A) & 0 &  \mathrm{c}(A) + i\frac{mc^{2} \delta t}{A} \mathrm{s}(A)
\end{bmatrix}
\end{eqnarray}
where we defined
\begin{eqnarray}
 \mathrm{c}(A) := \cos(A) \;\;, \;\; \mathrm{s}(A) := \sin(A)
\end{eqnarray}
and
\begin{eqnarray}
\label{eq:Afac}
 A := \sqrt{(mc^{2} \delta t)^{2} + \tilde{A}_{h,z}(j,k)^{2} + \tilde{A}_{h,r}(j,k)^{2} +\tilde{A}_{h,\theta}(j,k)^{2}} .
\end{eqnarray}
This ends the description of the numerical method.

\subsection{Higher order splitting}

The main numerical error in the scheme presented in the preceding section is due to the operator splitting and it can be shown that it is a first order method \cite{Lorin2011190} in time. This can be improved significantly by choosing higher order splitting schemes like the second order Strang-like splitting scheme given by (we omit the $(r,z)$ in arguments for notational convenience)
\begin{eqnarray}
\begin{array}{lll}
i\partial_{t} \psi^{(1)}(t) = \hat{A} \psi^{(1)}(t), & \psi^{(1)}(t_{n}) = \psi^n,  & t \in [t_{n},t_{n+\frac{1}{2}}) \nonumber \\
i\partial_{t} \psi^{(2)}(t) =\hat{B} \psi^{(2)}(t), & \psi^{(2)}(t_{n})  = \psi^{(1)}(t_{n+\frac{1}{2}}), & t \in [t_{n},t_{n+\frac{1}{2}}) \nonumber \\
i\partial_{t} \psi^{(3)}(t) = \hat{D} \psi^{(3)}(t), & \psi^{(3)}(t_{n}) = \psi^{(2)}(t_{n+\frac{1}{2}}),  & t \in [t_{n},t_{n+1}) \nonumber \\
i\partial_{t} \psi^{(4)}(t) = \hat{B} \psi^{(4)}(t), & \psi^{(4)}(t_{n+\frac{1}{2}}) = \psi^{(3)}(t_{n+1}), & t \in [t_{n+\frac{1}{2}},t_{n+1})  \nonumber \\
i\partial_{t} \psi^{(5)}(t) = \hat{A} \psi^{(5)}(t), & \psi^{(5)}(t_{n+\frac{1}{2}}) =  \psi^{(4)}(t_{n+1}),& t \in [t_{n+\frac{1}{2}},t_{n+1}) \nonumber \\
\end{array} \nonumber \\
\mbox{and} \;\; \psi^{n+1} = \psi^{(5)}(t_{n+1})
\end{eqnarray} 
Here, the time increments are $\delta t/2 = t_{n+\frac{1}{2}}-t_{n} = t_{n+1}-t_{n+\frac{1}{2}}$ for $\hat{A}$ and $\hat{B}$, and finally, $\delta t = t_{n+1} - t_{n} $ for $\hat{D}$.  This kind of splitting induces an error of $O(\delta t^{3})$. Then, the space discretization proceeds as in the lowest order splitting, leading to a second order numerical method \cite{Lorin2011190}. Of course, this can be improved to arbitrary order but this increases the computation time significantly. In this work, we consider only the lowest  order and second order splitting. 

For an axisymmetric system in 2-D or when the electromagnetic potential has no $z$-dependence, the Dirac equation becomes one dimensional. In this case, the last equation simplifies considerably because the number of steps in the splitting is reduced; it is possible to omit every step having the operator $\hat{B}$ (the same is true for the lowest order splitting in Eq. (\ref{eq:split1})).

\section{Numerical results}
\label{sec:num_results}

The numerical method exposed in the previous sections is tested in this section by considering some physical benchmarks. In the first case, the time evolution of Gaussian wave packets is investigated. This is one of the most simple systems and it allows analytical solutions in 2-D. The case of a 3-D wavepacket interacting with a laser field is also treated. The other system considered is the calculation of bound states of hydrogen-like atoms in 2-D and 3-D. In this case, the eigenenergy is calculated by using a standard method in non-relativistic physics: the Feit-Fleck spectral method \cite{Feit1982412}. The latter allows the computation of eigenpairs for a given static potential from the solution of the time-dependent equation. The eigenenergies obtained in this way are compared to analytical results. 

All the computations are performed in atomic units where the electron mass is given by $m=1.0$ and the speed of light is $c = \alpha^{-1}$ where $\alpha$ is the fine structure constant given by $\alpha \approx 1/137.0359895$. Note that in these units, we also have the electron charge $|e|=1$ and the Plank constant $\hbar = 1.0$. Also, we use Dirichlet's boundary conditions at $r=r_{\rm max}$, $z=z_{\rm min}$ and $z_{\rm max}$, that is
\begin{eqnarray}
\partial \Omega_{r=r_{\rm max},z,\theta} &=& 0 ,\\
\partial \Omega_{r,z=z_{\rm min},\theta} &=& 0, \\
\partial \Omega_{r,z=z_{\rm max},\theta} &=& 0 .
\end{eqnarray}
The value of $r_{\rm max},z_{\rm min}$ and $z_{\rm max}$ are chosen large enough such that no spurious reflection occurs. Transparent boundary conditions are presently under study and will be the main topic of a future publication.

%
%

\subsection{Gaussian wave packets and the order of convergence of the radial part}

Gaussian wave packets are certainly among the most elementary systems that can be studied with the Dirac equation as they are built from the superposition of plane waves in vacuum (when there is no external field). Physically, they represent electrons localized in space, so they have a significant importance in many applications. For this reason, they have been studied in many different circumstances \cite{PhysRevA.59.604,Mocken2008868,Mocken2004558,Su:98,PhysRevA.63.032107,2004quant.ph..9079T,PhysRevA.82.052115,FillionGourdeau2012}. In this study, the time evolution of a simple wave packet will be used to study the convergence and to validate the numerical method. Thus, the initial condition considered is given by
\begin{equation}
\label{eq:Gauss_in}
 \psi(t=0, r) = \mathcal{N} \begin{bmatrix} r^{|\mu_{1}|} \\ 0 \\ 0 \\ 0 \end{bmatrix} e^{- \frac{r^{2}}{4 \Delta^{2}}}
\end{equation}
where $\mathcal{N}$ is a normalization constant and $\Delta$ characterizes the Gaussian width. For $\mu_{1} = 0$, this wave packet represents a spin-up massive electron. The shape of this initial condition is chosen such that conditions in Eqs. \eqref{eq:exp_form1} to \eqref{eq:exp_form4} are fulfilled at small $r$ while preserving axial symmetry. Then, the numerical scheme described in previous sections can be used to evaluate the time evolution of this wave packet. 

When $\mu_{1} = 0$, an analytical solution can be calculated (up to a numerical integration) by Fourier transform methods, as shown in \ref{app:sol_2-D_WP}. To make a comparison with the numerical results, the remaining integrals are calculated in Maple using a numerical scheme based on an adaptive Gauss-Kronrod quadrature (with Gauss 30-point and Kronrod 61-point rules), well suited for oscillatory integrands.  

\begin{figure}
\centering
\includegraphics[width=0.6\textwidth]{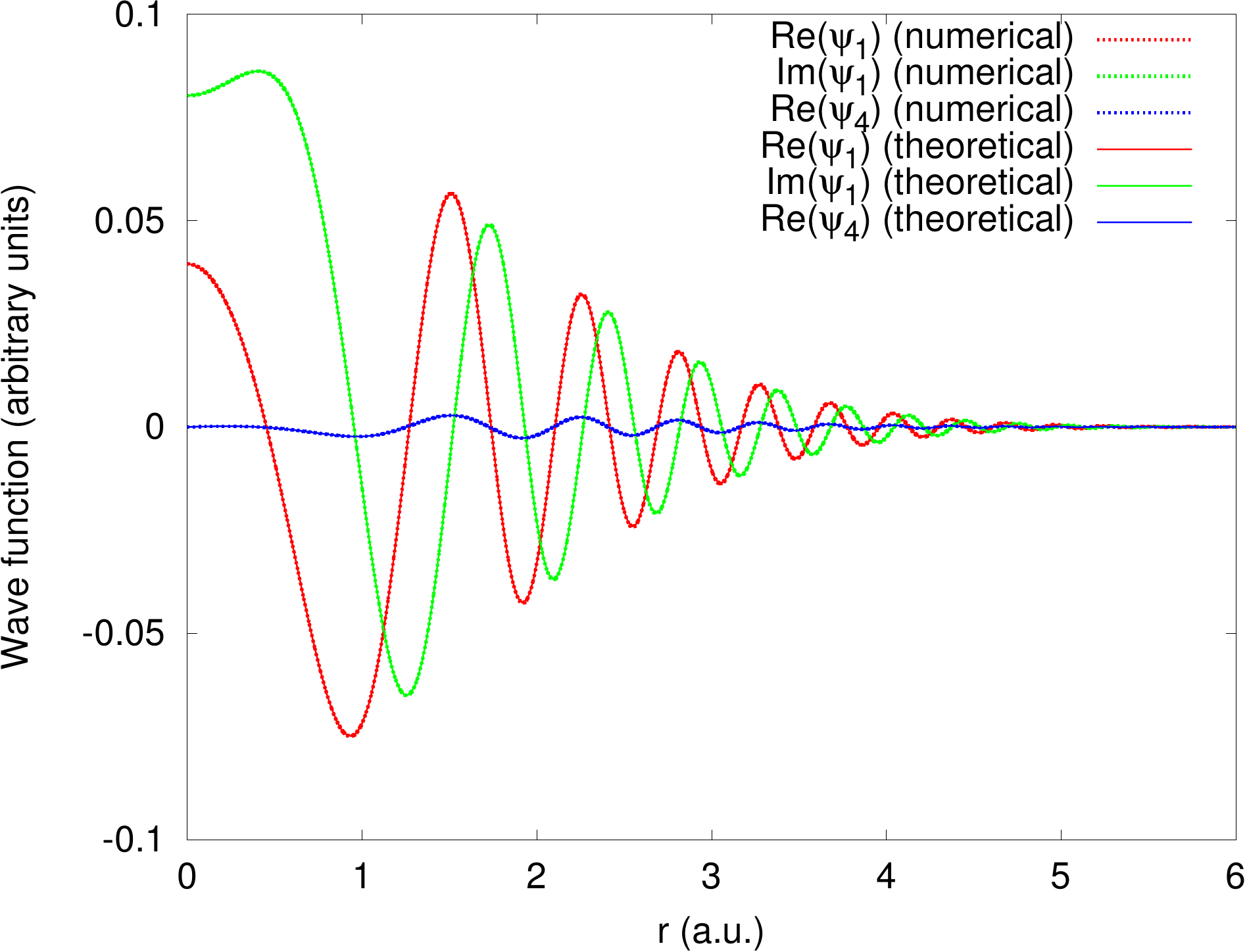}
\caption{Results for the non-zero components of the Gaussian wave packet with an initial width of $\Delta = 0.1$, evaluated at time $t=0.143$ a.u.. The theoretical and calculated curves overlap. }
\label{fig:gaussian_d0p1}
\end{figure}

\begin{figure}
\centering
\includegraphics[width=0.6\textwidth]{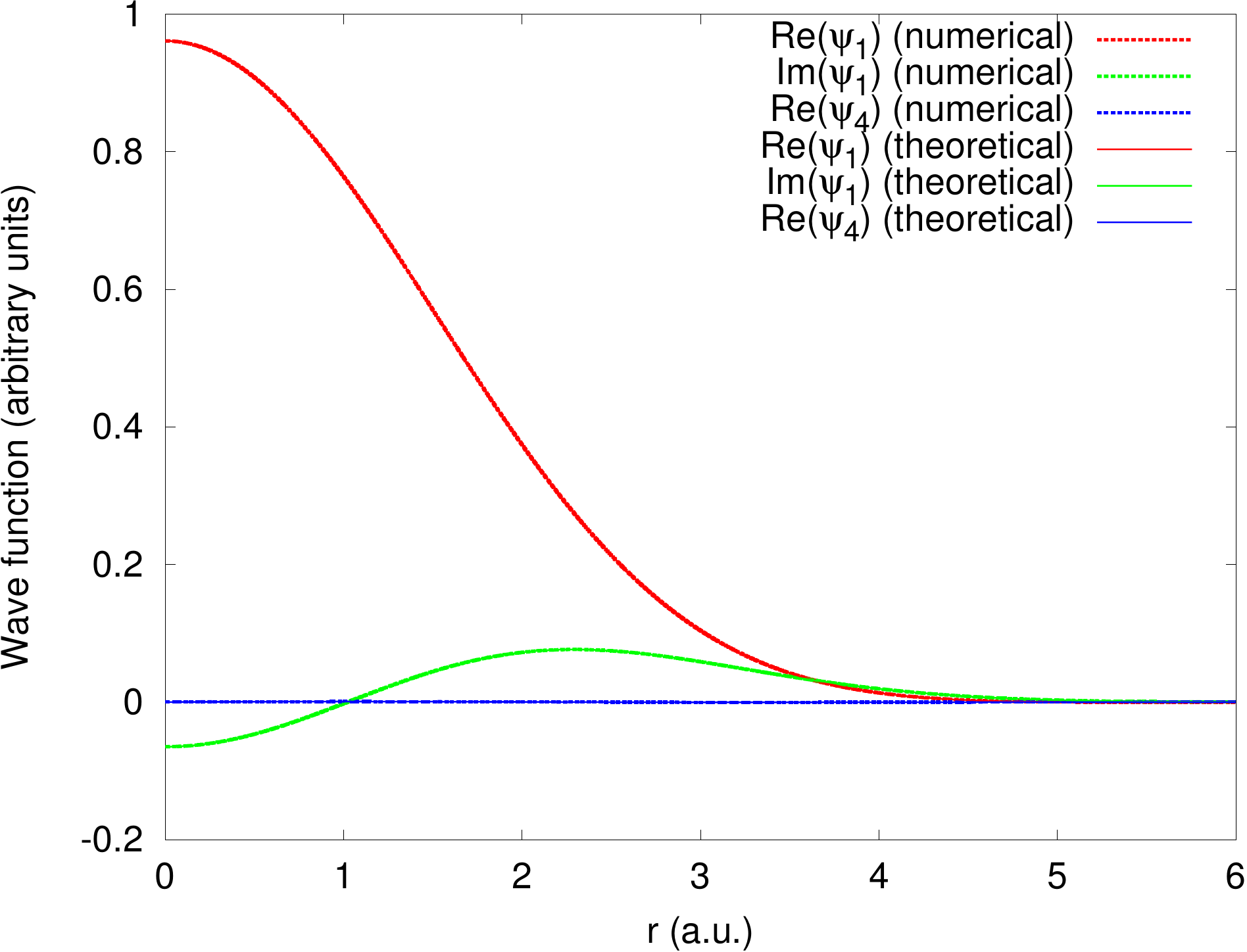}
\caption{Results for the non-zero components of the Gaussian wave packet with an initial width of $\Delta = 1.0$, evaluated at time $t=0.570$ a.u.. The theoretical and calculated curves overlap. }
\label{fig:gaussian_d1}
\end{figure}

The numerical results are obtained from the scheme described in previous sections, using the first order splitting. The width of the wave packet is set to $0.1$ and $1.0$ a.u. while the domain external boundary is $r_{\rm max} = 10.$ a.u., respectively. The time step in both cases is $\delta t = 1.113 \times 10^{-6}$ a.u.. The non-zero components of the wave function are displayed in Figs. \ref{fig:gaussian_d0p1} and \ref{fig:gaussian_d1} at a time of 0.223 a.u. and 0.557 a.u., respectively. In both cases, the calculated results are in very good agreement with the theoretical ones. Also, there are no spurious oscillations close to the boundary $r=0$, as was observed in \cite{0022-3700-16-11-017} for a finite difference discretization.

Then, the order of convergence of the method in the radial direction is investigated. For the longitudinal direction, the numerical scheme is the same as the Cartesian coordinates case, and this was treated in \cite{Lorin2011190}. Therefore, the focus here is on the radial operator.

\begin{figure}
\centering
\includegraphics[width=0.6\textwidth]{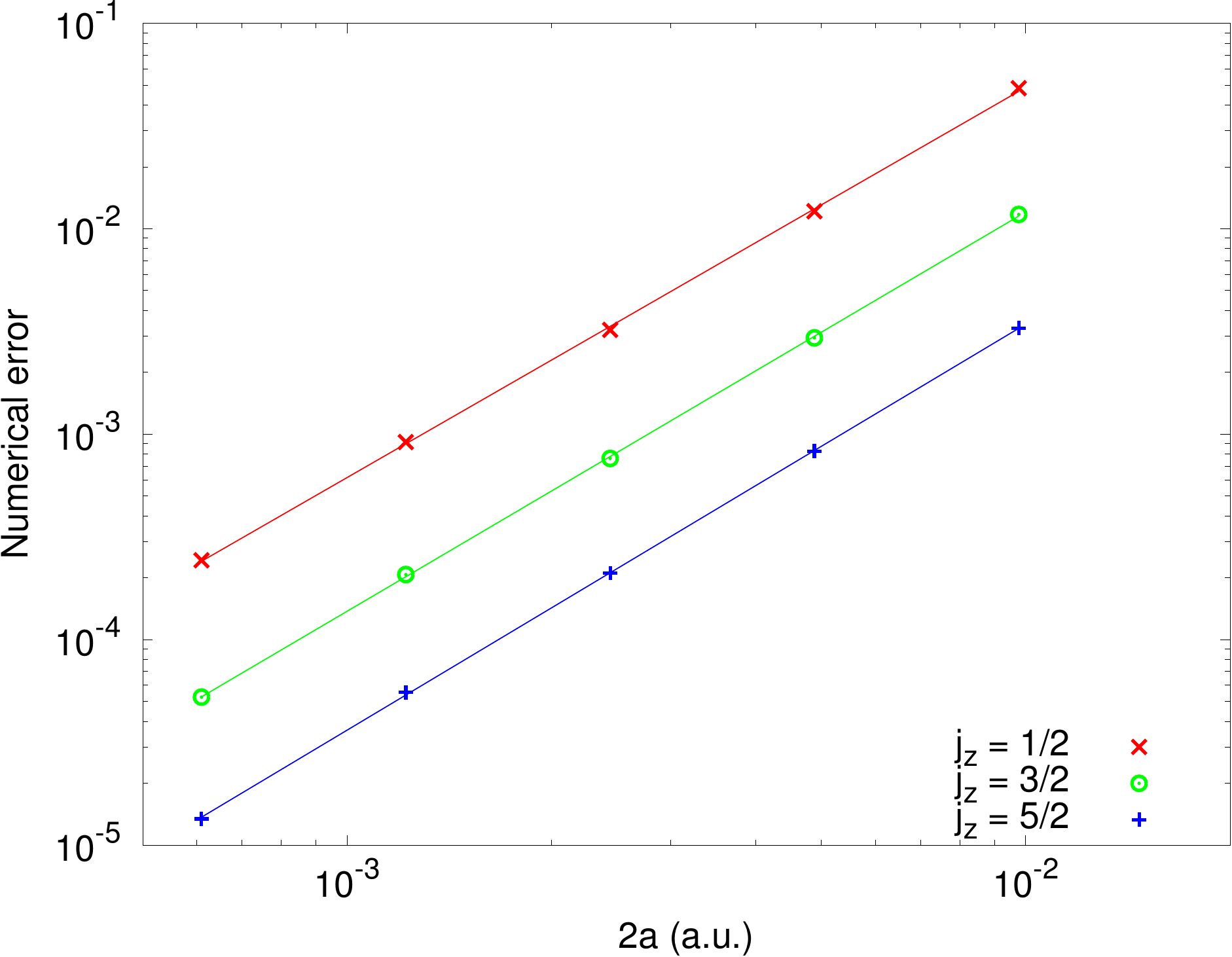}
\caption{Numerical error in the $\ell^{2}$-norm for a wave packet in the radial direction evaluated at time $t=0.0356$ a.u., for different mesh sizes. The lines are linear fits of these errors and the slope gives the order of convergence $q$. }
\label{fig:order_conv}
\end{figure}

The methodology used to determine the order of convergence $q$ is based on the evaluation of the numerical error in the $\ell^{2}$ norm defined by 
\begin{eqnarray}
\label{eq:num_erro}
E_{\ell^{2}}(t) := \norm{\psi_{\rm exact}(t) - \psi(t)}_{\ell^{2}},
\end{eqnarray}  
where $\psi_{\rm exact}$ is the exact solution. The latter is approximated by the solution calculated on the mesh with the smallest space step (converged solution). The numerical error is calculated for different mesh sizes and angular momenta. The results are shown in Fig. \ref{fig:order_conv} on a logarithmic plot. The slope of the data can be easily found by linear interpolation and gives the order of convergence. The numerical values of $q$ are given in Table \ref{table:order_conv} for the considered angular momenta. For all values of $j_{z}$, the order of convergence in the radial direction is close the order of convergence in the longitudinal direction ($q \approx 2.0$ in the $z$-coordinate \cite{Lorin2011190}).

\begin{table}[h]
\caption{Results for the order of convergence $q$ determined numerically}
\centering
\begin{tabular}{cc}
 \hline \hline
 $j_{z}$ & Order of convergence ($q$)\\
 \hline  
$\frac{1}{2}$& 1.8999 \\
$\frac{3}{2}$&  1.9435\\
$\frac{5}{2}$&  1.9767\\
 \hline \hline 
\end{tabular} 
\label{table:order_conv}
\end{table}

\subsection{3-D Gaussian wave packet in a counterpropagating laser field}
\label{sec:3d_gaussian}
One of the important applications of our numerical method concerns the interaction of matter with very high intensity lasers. In the last few decades, the laser intensities reached in laboratories have increased significantly such that it is now plausible to start observing relativistic and Quantum Electrodynamic (QED) effects \cite{Salamin200641}. One of the most important QED observables is the spontaneous production of electron-positron pairs from the laser field. Studying this effect from the theoretical point of view requires a solution of the time-dependent Dirac equation \cite{PhysRevC.71.024904,PhysRevLett.102.080402}. Therefore, the next example considered is that of a wave packet in a time-varying homogeneous electric field. Physically, it corresponds to the interaction of an electron with a counterpropagating laser. More specifically, the electric field is given by $E(t)=E_{0}f(t)\cos(\omega t)$ where $E_{0}$ is the electric field strength, $\omega$ is the laser frequency and the envelope function is given by
\begin{eqnarray}
\label{eq:pot_laser}
f(t) = 
\begin{cases}
 \frac{\omega}{n\pi} t & \mbox{for} \;\; t\in [0, n\pi/\omega], \\
1 & \mbox{for} \;\; t\in [n\pi/\omega, (n+n')\pi/\omega], \\
-\frac{\omega}{n\pi}\left(t- \frac{(2n+n')\pi}{\omega}\right) & \mbox{for} \;\; t\in [(n+n')\pi/\omega, (2n+n')\pi/\omega] ,\\
\end{cases}
\end{eqnarray}
where $n,n' \in \mathbb{N}$ counts the number of half-cycles for linear ramping and for a constant enveloppe, respectively. Note that there is no space dependence in $E(t)$, which corresponds to the field of the laser at the anti-node of the standing wave. We work in a gauge where $A_{0}=0$ and thus, the electric vector potential is given by $A_{x}(t)=\int^{t} dt' E(t')$. The initial state represents a positive energy wave packet at rest and is assumed to be given by Eq. \eqref{eq:Gauss_in} with $\mu_{1}=0$ (we consider $j_{z}=1/2$). 

\begin{figure}
\centering
\subfloat[]{\includegraphics[width=0.8\textwidth]{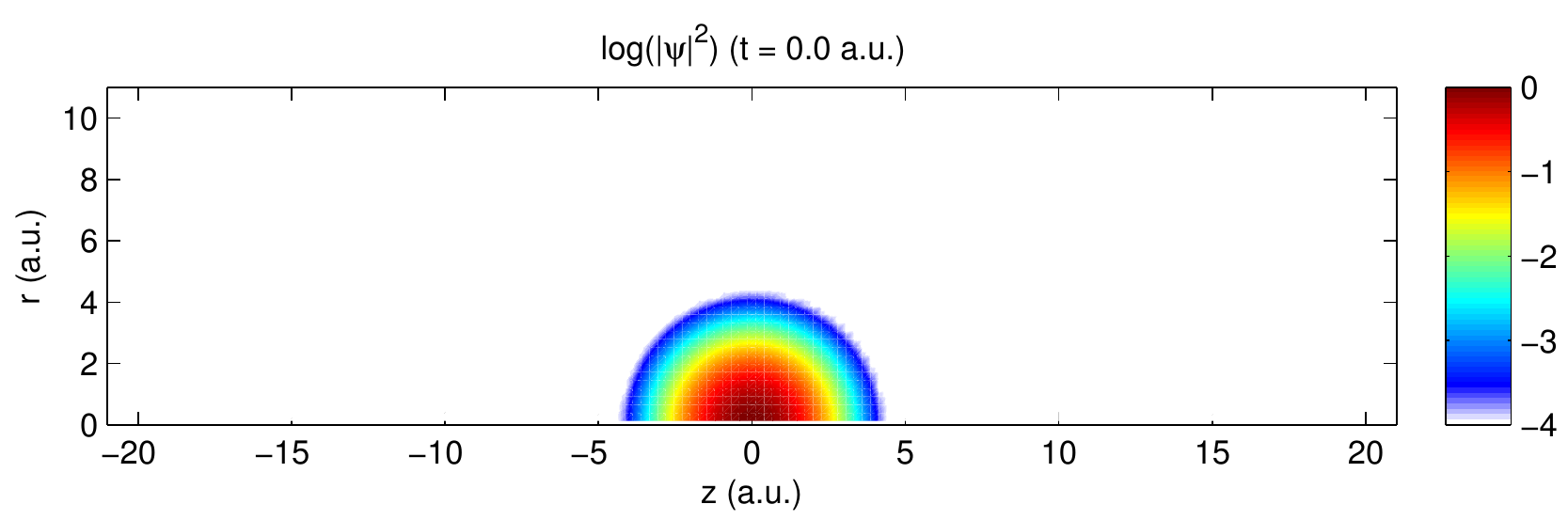}} \\
\subfloat[]{\includegraphics[width=0.8\textwidth]{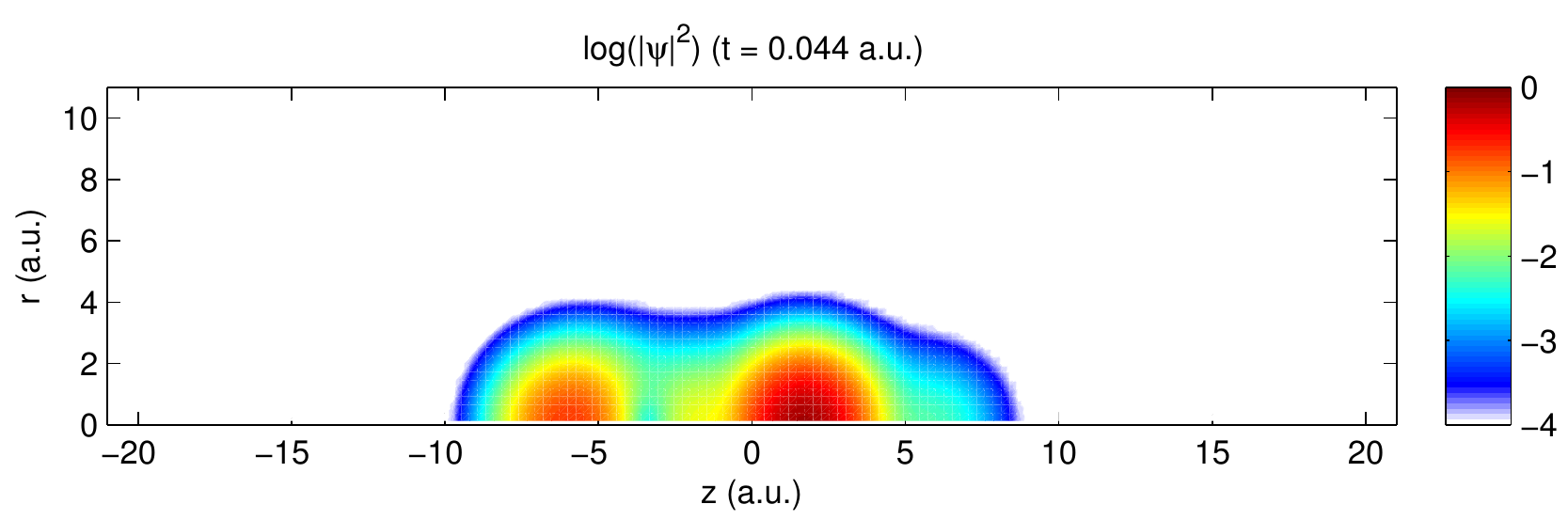}} \\
\subfloat[]{\includegraphics[width=0.8\textwidth]{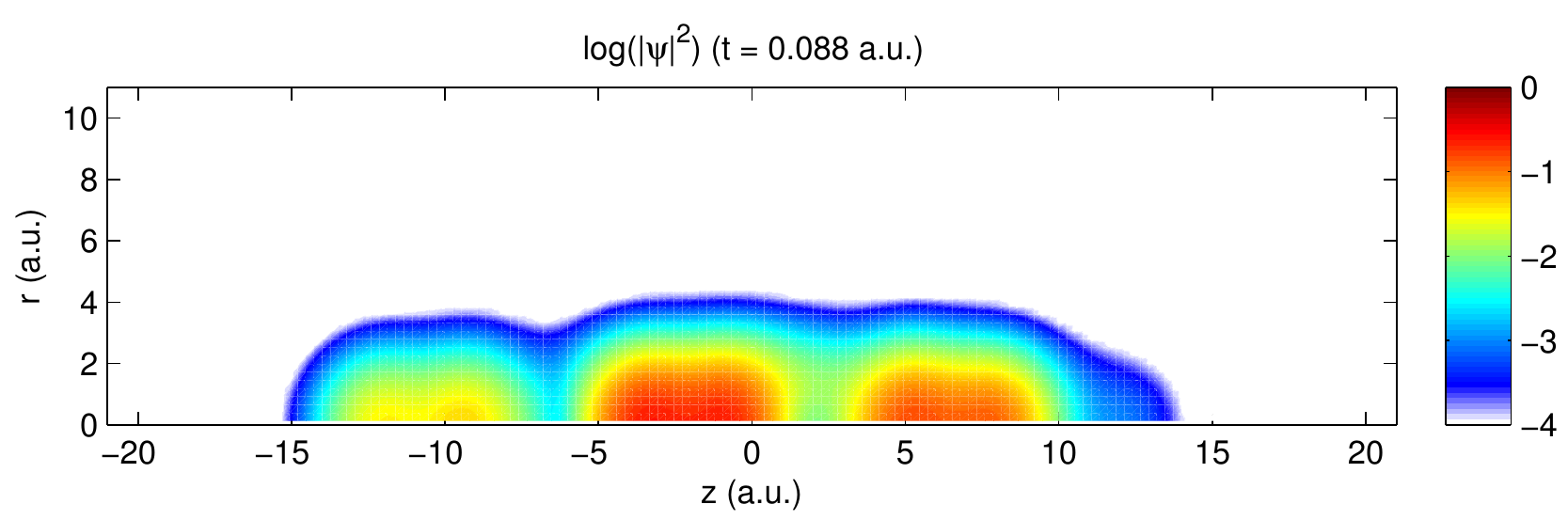}} \\
\subfloat[]{\includegraphics[width=0.8\textwidth]{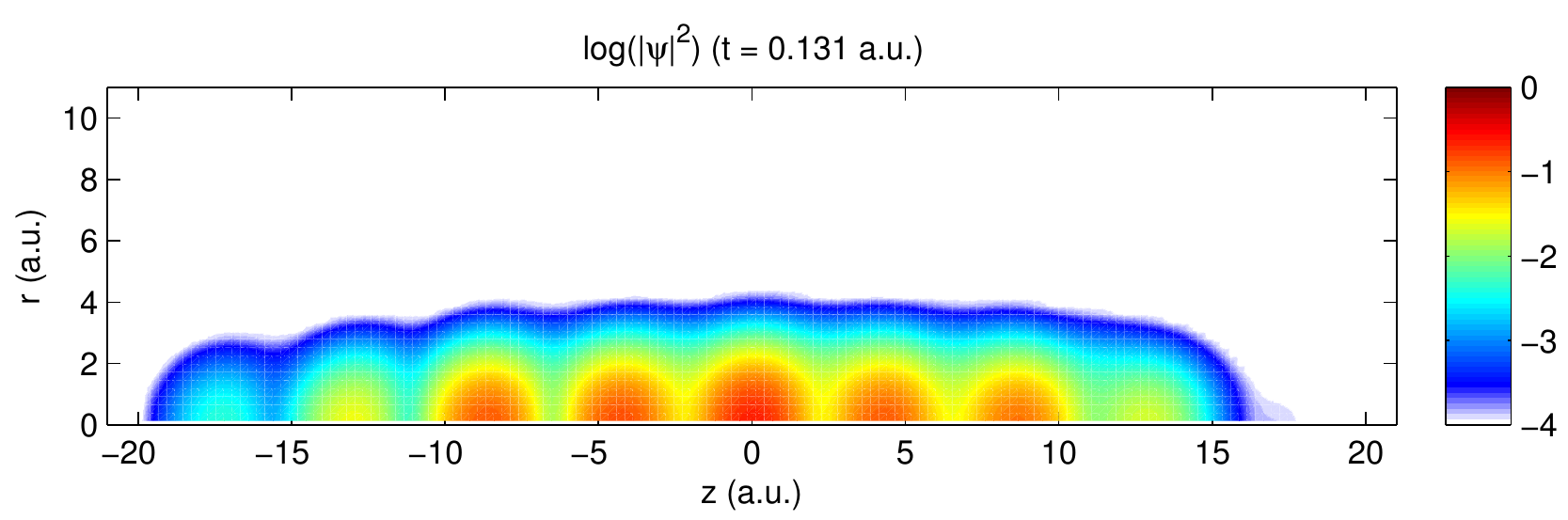}} 
\caption{Time evolution of the density $|\psi|^{2}$ for a wave packet in a counterpropagating laser field at (a) $t=0.0$ a.u., (b) $t=0.044$ a.u., (c) $F=0.088$ a.u. and (d) $t=0.131$ a.u..}
\label{fig:gaussian_laser}
\end{figure}

The simulation is shown in Fig. \ref{fig:gaussian_laser} for $\omega = 100$ a.u. and $E_{0} = 3.65 \times 10^{6}$ a.u., which is slightly higher than Schwinger's critical field ($2.6 \times 10^{6}$ a.u.) at which static electron-positron pair production starts to be important. The width of the initial wave packet is set to $1.0$ a.u. while the domain external boundaries are at $r_{\rm max} = 10.$ a.u., $z_{\rm min}=-20.$ a.u. and $z_{\rm max} = 20.$ a.u.. The time step is $\delta t = 4.86 \times 10^{-5}$ a.u..

The numerical results show that secondary peaks are formed at each cycles. The group velocities of these peaks and that of the primary peaks are in opposite direction and thus, they are carrying different electric charge. This can then be interpreted as the production of electron-positron pairs \cite{PhysRevLett.102.080402,Mocken2004558}. This is because we started with an initial state containing only positive energy states. By interacting with the intense electromagnetic field, negative energy states are created. It should be noted however that a full calculation of the pair production rate requires a summation over all the positive and negative energy states \cite{PhysRevLett.102.080402}: this is not performed here.

\subsection{Bound states}

It is very important in many applications to be able to evaluate bound states of the Dirac operator because many observables can be related to these entities. Also, the static bound states can serve as initial conditions for the dynamical evolution of the system under study. For these reasons, many numerical methods were developed to compute the Dirac eigenenergies and eigenpairs for a given static potential \cite{Kullie2004215,esteban2007,Dolbeault2003,Desclaux2003453,PhysRevA.85.022506,PhysRevLett.93.130405,PhysRevA.62.022508}. It is possible to compute time-independent wave functions from a time-dependent numerical scheme by using the well-known Feit-Fleck method \cite{Feit1982412}, as explained in the following.

\subsubsection{Spectral method}
\label{sec:spec_meth}
In this section, the spectral method used to compute the eigenenergies is presented. Most of this section is based on \cite{Feit1982412} where this method was used to evaluate the eigenstates of the Schr\"odinger equation and where more details can be found. This method was also used for the Dirac equation in \cite{Mocken2004558,Mocken2008868,PhysRevA.83.063414,Bauke2011} to calculate the eigenfunction for hydrogen-like atoms. 

The main ingredient of this numerical scheme is the auto-correlation function defined by
\begin{eqnarray}
C(t) &:=& \langle \Psi(0) | \Psi(t)\rangle ,\\
&:=& \int_{\mathbb{R}^{3}} d^{3}\mathbf{x} \Psi^{\dagger}(0,\mathbf{x})\Psi(t,\mathbf{x}).
\end{eqnarray}
where $\Psi(0,\mathbf{x})$ is an arbitrary trial function. When the potential is static (independent of time), the wave function can be expressed as a superposition of eigenstates. It can then be easily demonstrated that the Fourier transform of the auto-correlation function $\widehat{C}(E)$ is sharply peaked at the eigenenergy values;  $\widehat{C}(E)$ becomes a sum over Dirac delta functions positioned at the exact bound state energies. This however implies that the wave function should be evolved to an infinite time to evaluate the Fourier transform, which of course, is impossible in a numerical calculation. 

The same procedure can however be performed when only a finite time is available. In this case, the Fourier transform is calculated as
\begin{eqnarray}
\widehat{C}(E) \sim \int_{-\infty}^{\infty}dt w(t)C(t)e^{iEt}
\end{eqnarray}
where $w(t)$ is a window function. The latter allows to determine the functional form of the lineshape, i.e. the Fourier transform of the window function gives the equation of the eigenenergy peaks. It is convenient to choose the Hann window function given by
\begin{eqnarray}
w(t) = 
\begin{cases}
\frac{1-\cos \left(\frac{2\pi t}{T}\right)}{T} & \mbox{for} \; t \in [0,T] \\
0 & \mbox{for} \; t \in (-\infty,0)\cup (T,\infty)
\end{cases} ,
\end{eqnarray}
where $T$ is the final time of the calculation. This choice of window function allows an accurate determination of resonance position because the resulting lineshape has no side lobes which could be confounded with other resonances. 

As seen previously, the eigenenergies can be determined by looking at the power spectrum of the auto-correlation function. Once the energy $E$ of the bound state is determined, the corresponding eigenstate $\Psi_{E}$ can be calculated by using
\begin{eqnarray}
\label{eq:eigen_state}
\Psi_{E}(\mathbf{x}) = \int_{0}^{T} dt \Psi(t,\mathbf{x}) w(t)e^{iEt},
\end{eqnarray}
where $\Psi(0,\mathbf{x})$ is an arbitrary trial function.

The numerical procedure can be summarized as follow:
\begin{enumerate}
	\item Determine the eigenenergy
\begin{itemize}
	\item Choose a trial function.
	\item Evolve the trial function in time by computing $C(t)$ at each time step.
	\item Compute the Fast Fourier Transform (FFT) of $C(t)w(t)$ to obtain the power spectrum.
\end{itemize}
	\item Determine the eigenfunction.
\begin{itemize}
	\item Choose a trial function
	\item Set the value of the energy to the one determined in the previous step.
	\item Evolve the trial function in time by calculating Eq. (\ref{eq:eigen_state}) at each time step.
\end{itemize}
\end{enumerate}
This numerical method requires at least two calculations of the time-dependent Dirac equation to obtain the eigenstate. This is clearly not as efficient as other methods such as variational schemes \cite{QUA:QUA560250112,Dolbeault2003,PhysRevA.85.022506}. However, it is very easy to implement and it allows to evaluate the eigenstate of the grid used in the split-step method; in contradistinction with basis set expansion methods which can not be easily adapted to our numerical method.

\subsubsection{1-D exponential potential}

By combining the Feit-Fleck method with the scheme for solving the time-dependent Dirac equation, it is possible to evaluate the spectrum and eigenfunction of any bounding potentials consistent with the chosen grid. The first test of the spectral method is for a simple exponential potential in 1-D. We consider a static scalar potential given by
\begin{eqnarray}
 V(z) = -\frac{g}{2a} e^{-\frac{|z|}{a}}
\end{eqnarray}
where $g$ and $a$ are parameters characterizing its strength and width, respectively. The Dirac equation can be solved exactly in this case and the eigenenergies $E$ are solutions of (in units where $c=1$)\cite{DominguezAdame1995275}
\begin{eqnarray}
\left(\frac{E-iq}{m} \right)^{2} = \frac{_{1}F_{1}(qa+iEa,1+2qa;ig) _{1}F_{1}(1+qa-iEa,1+2qa;-ig)}{_{1}F_{1}(qa-iEa,1+2qa;-ig) _{1}F_{1}(1+qa+iEa,1+2qa;ig)}
\end{eqnarray}
where $q = \sqrt{m^{2}-E^{2}}$ and $_{1}F_{1}$ is the confluent hypergeometric function. This transcendental equation can be solved numerically and thus,  can be compared to the numerical results obtained from the spectral method. 

\begin{figure}
\centering
\includegraphics[width=0.7\textwidth]{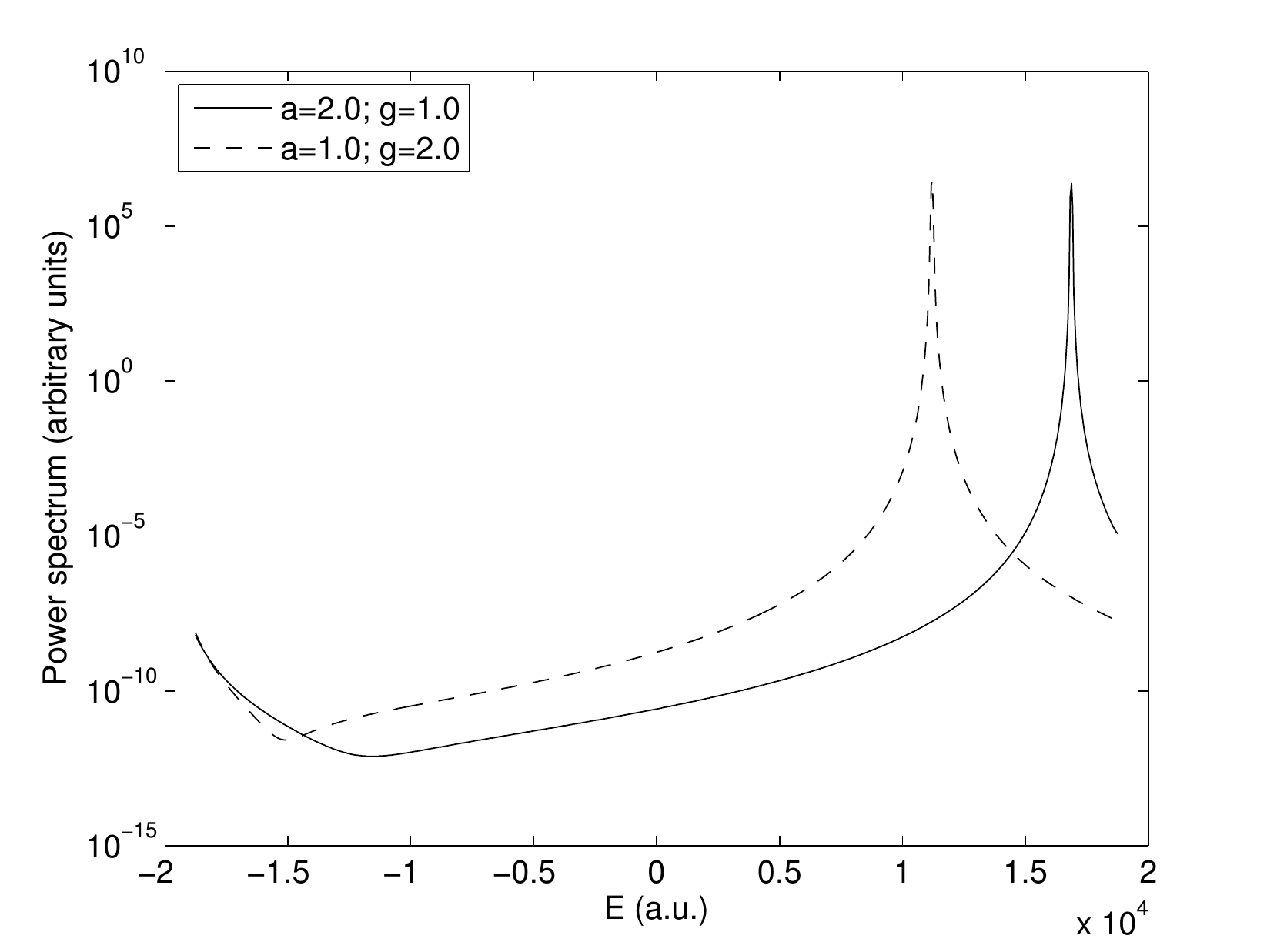}
\caption{Power spectrum for the exponential potential for two cases: $g=1.0;\;a=2.0$ and $g=2.0;\;a=1.0$. The peak is at the position of the ground state energy. The analytical solution yields ground states of $E=16866.06$ a.u. and $E=11201.44$ a.u., respectively, so the peaks are at the right position.  }
\label{fig:spec_exp}
\end{figure}

For calculation using the Feit-Fleck method, we start with a trial function given by a gaussian wave packet representing a spin up static electron (see Eq. \eqref{eq:psi_init_2-DWP}, but set $x=z$ and $y=0$). We consider a 1-D domain, the $z$-axis, and make $2\times 10^{6}$ time iterations to get the required precision. The power spectra obtained for two cases ($g=1.0;\;a=2.0$ and $g=2.0;\;a=1.0$) are shown in Fig. \ref{fig:spec_exp}. We find ground state energies of $E_{\rm ground} = 16875$ and $E_{\rm ground} = 11218$, for the two cases, respectively, which is close to the analytical solution, where we get 
$E=16866.06$ a.u. and $E=11201.44$ (the relative difference is less than 0.15\%)

\subsubsection{3-D hydrogen-like atom in a laser field}

This last example concerns the interaction of a laser field with the ground state electron of a hydrogen-like atom (1$s$ orbital). The ion is modelled by the 3-D Coulomb potential. The latter can be solved analytically, when no laser field is present, and represents physically the electric potential of a point charge. Numerically however, it is problematic because the wave function and potential have a singularity at the charge position, in $r=0$. Thus, it does not obey some of the assumptions required by the numerical method presented in this article. For this reason, rather than using the Coulomb potential directly, a regularized potential is used. It is given by
\begin{eqnarray}
V(r) = 
\begin{cases}
\frac{Z}{2R} \left(\frac{r^{2}}{R^{2}} - 3  \right) & \mbox{for} \; r<R \\
-\frac{Z}{r}  & \mbox{for} \; r \geq R
\end{cases}
\end{eqnarray}
where $R$ here is the radius of the nucleus and $Z$ is the atomic number. For $r<R$, this potential corresponds to the electric field of a charged sphere having a radius $R$ with constant distribution of charge. In this work, this is chosen to insure that $\psi \in C^{\infty}$ such that relevant numerical results can be obtained. In the following calculation, the radius of the nucleus is set to $R = 0.01$ a.u. while the atomic number is $Z=10$.  

\begin{figure}
\centering
\includegraphics[width=0.6\textwidth]{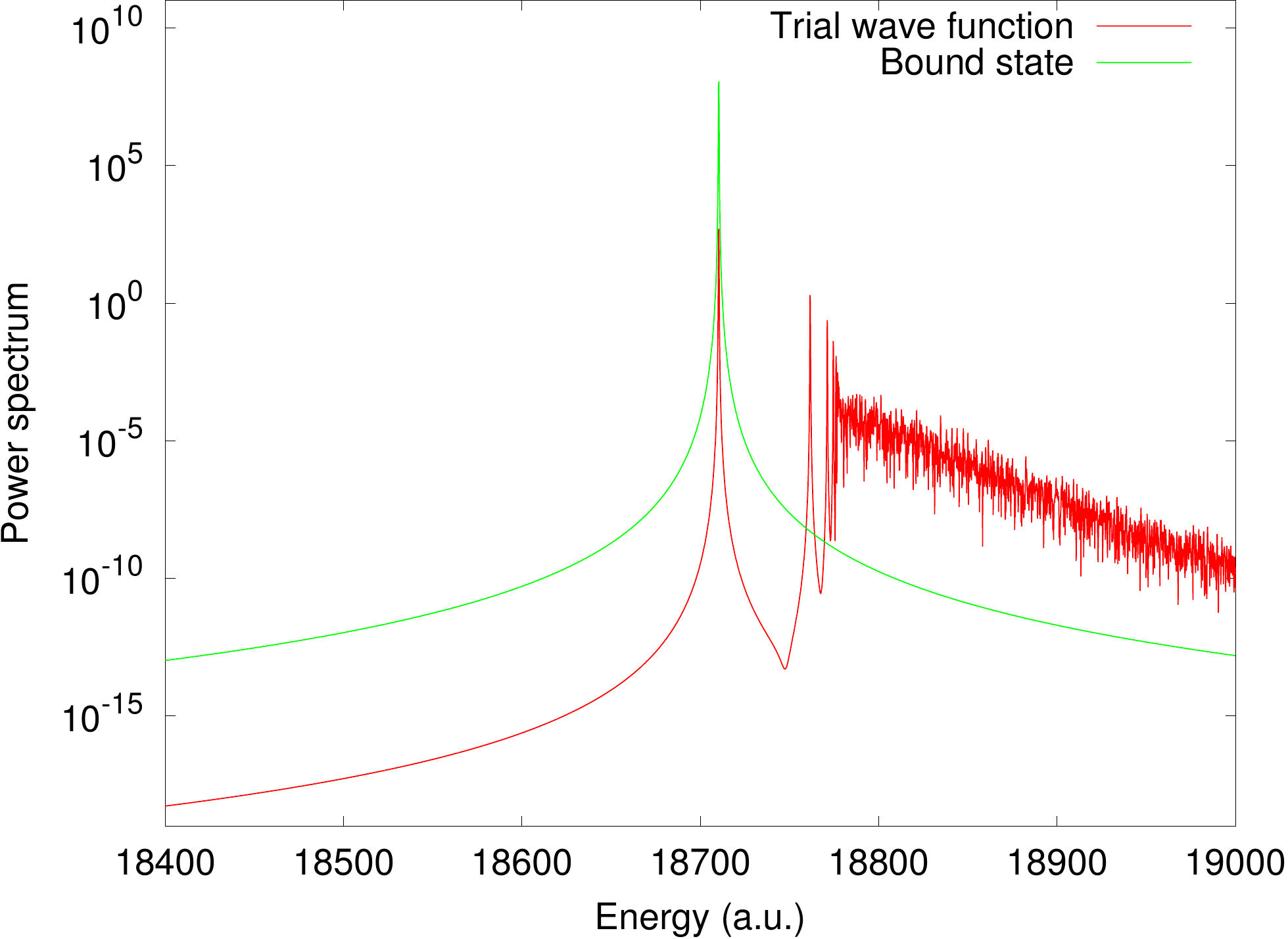}
\caption{Results for the power spectrum of trial wave function and the bound state constructed from the Feit-Fleck method. The Feit-Fleck method filters the wave function such that only the bound state remains.}
\label{fig:power_spec}
\end{figure}

\begin{figure}
\centering
\begin{overpic}[scale=0.7,unit=1\textwidth]
{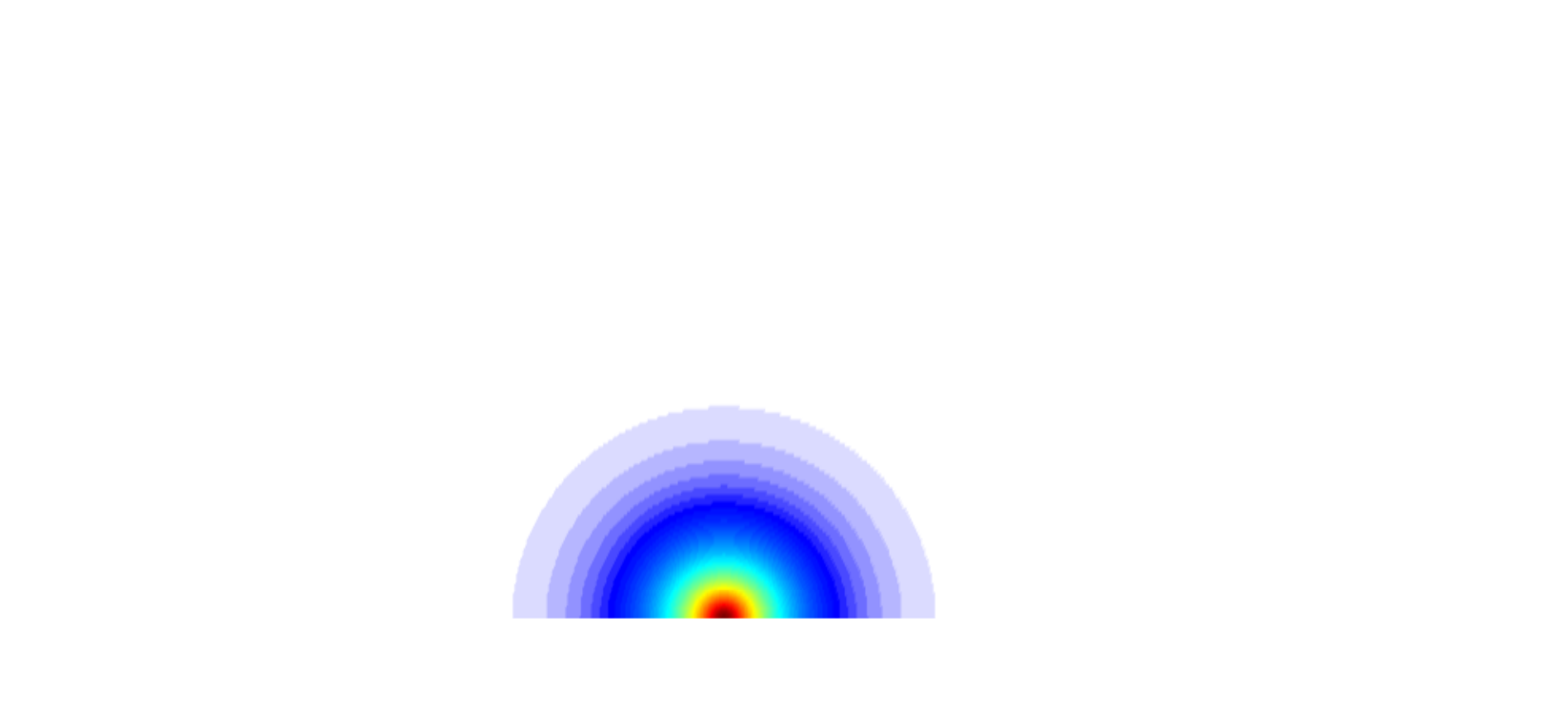}
\put(0,0){\includegraphics[scale=0.7]%
{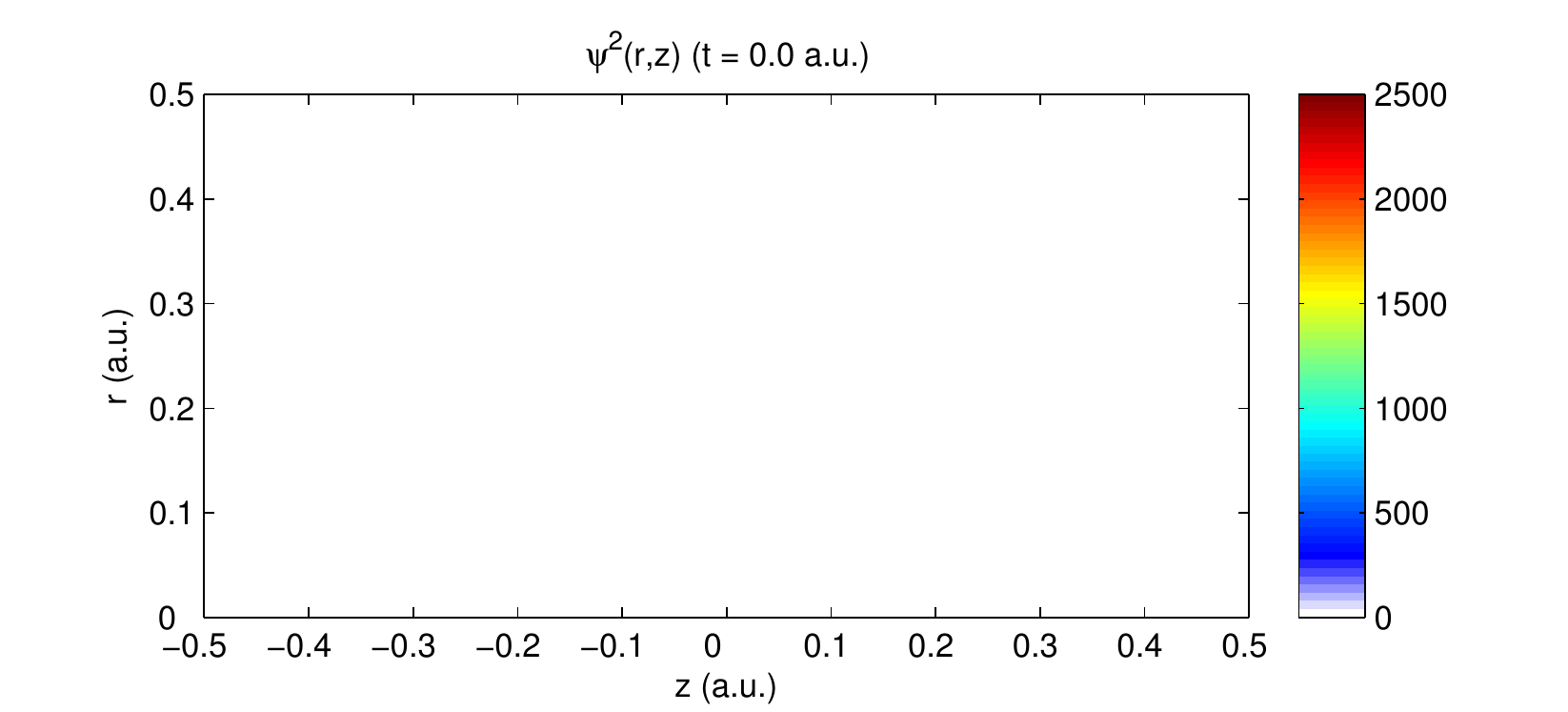}}
\end{overpic}
\caption{Ground state of a hydrogen-like atom with $Z=10$.}
\label{fig:ground_state}
\end{figure}

The initial state is prepared by using the spectral method presented in Sec. \ref{sec:spec_meth}. The trial state is a wave packet centred on the atom, it is given by
\begin{equation}
\label{eq:Gauss_in_atom}
 \psi(t=0, r,z) = \mathcal{N} \begin{bmatrix} r^{|\mu_{1}|} \\ 0 \\ 0 \\ 0 \end{bmatrix} e^{- \frac{r^{2}+z^{2}}{4 \Delta^{2}}}.
\end{equation} 
In the first step of the calculation, where the spectrum is evaluated, this initial state is evolved up to $t_{f} = 47.5$ a.u., making for an energy resolution of $\delta E \approx 0.13$ a.u. in the spectral method. It was then determined, from the power spectrum of the trial wave function shown in Fig. \ref{fig:power_spec}, that the ground state energy is $E_{\rm ground} \approx 18710.3$ a.u.. This value is close to the analytical Coulomb ground state energy given by $E_{\rm ground} \approx 18729.9$, with a relative difference of $\delta_{\rm rel} \approx 0.1 \%$. The ground state is then constructed from the same trial function by using Eq. \eqref{eq:eigen_state}. The spectrum of this bound state is also evaluated as a consistency check and is depicted in Fig. \ref{fig:power_spec}. A comparison of the trial state and bound state spectra demonstrates clearly that the spectral method filters out the unwanted frequencies while keeping only the ground state component. The resulting ground state is shown in Fig. \ref{fig:ground_state}.

\begin{figure}
\centering
\begin{overpic}[scale=1.0,unit=1\textwidth]
{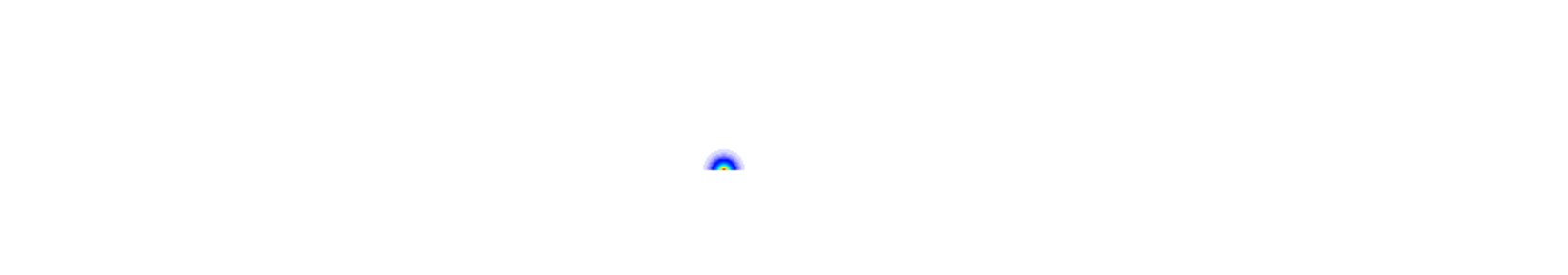}
\put(0,0){\includegraphics[scale=1.0]%
{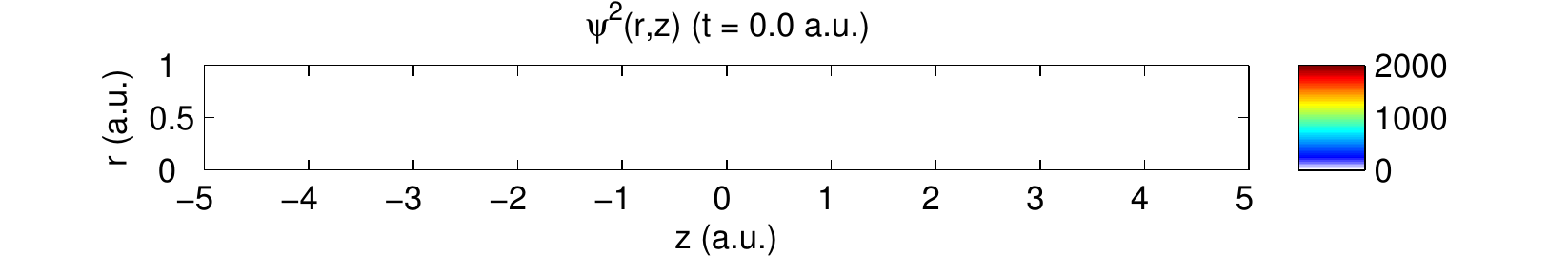}}
\end{overpic} \\ 
\begin{overpic}[scale=1.0,unit=1\textwidth]
{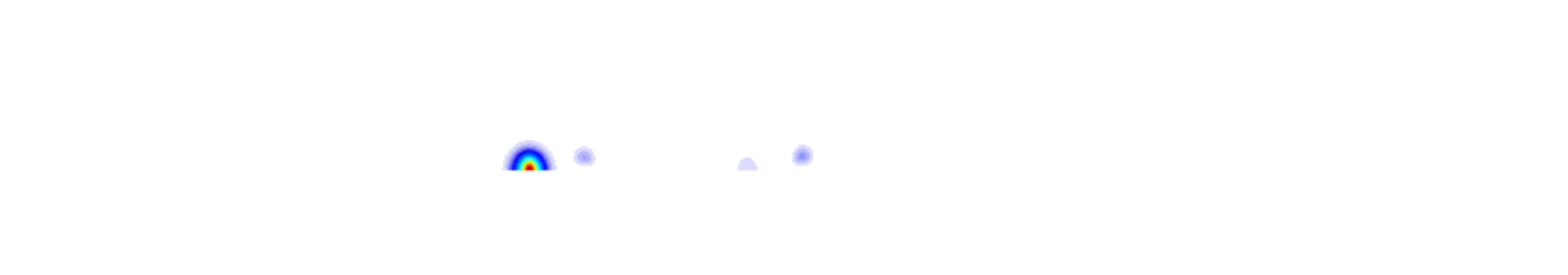}
\put(0,0){\includegraphics[scale=1.0]%
{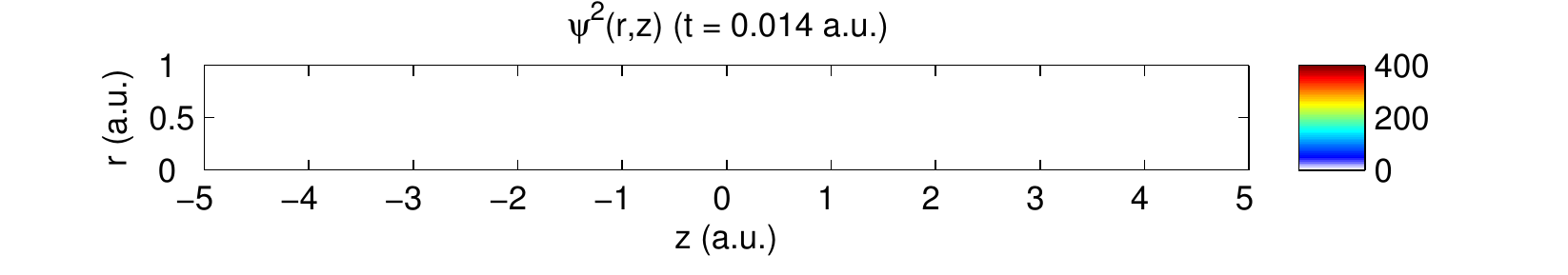}}
\end{overpic} \\ 
\begin{overpic}[scale=1.0,unit=1\textwidth]
{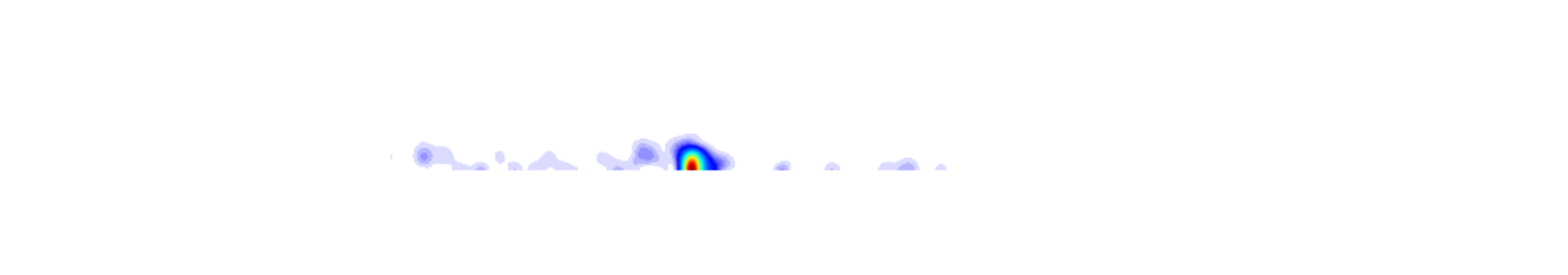}
\put(0,0){\includegraphics[scale=1.0]%
{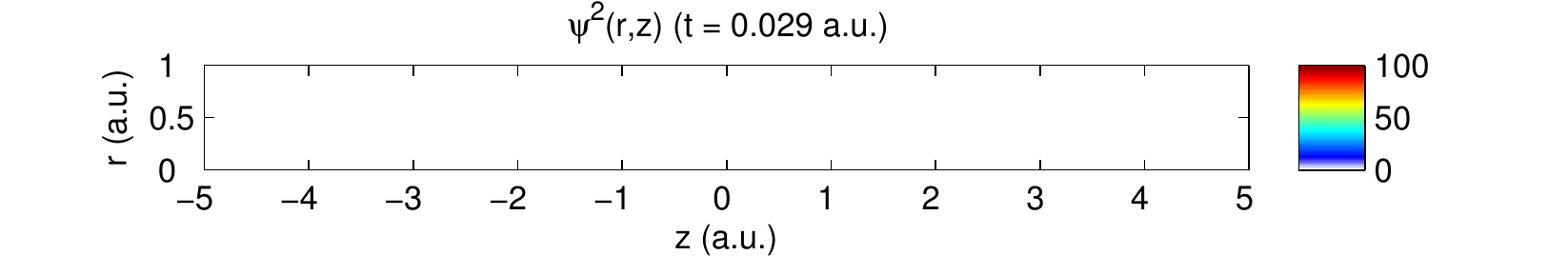}}
\end{overpic} \\ 
\begin{overpic}[scale=1.0,unit=1\textwidth]
{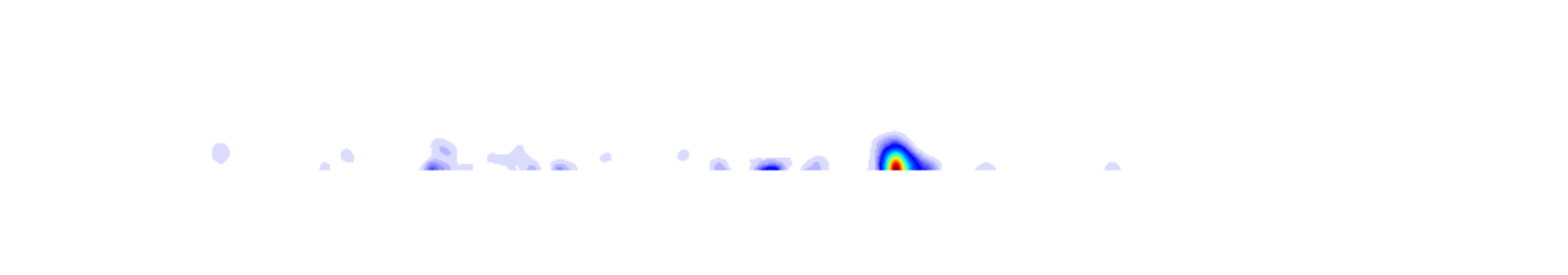}
\put(0,0){\includegraphics[scale=1.0]%
{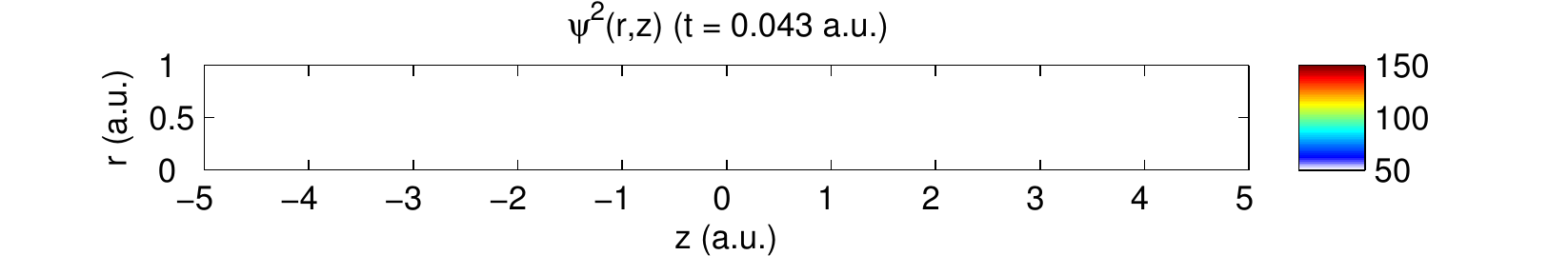}}
\end{overpic} \\
\begin{overpic}[scale=1.0,unit=1\textwidth]
{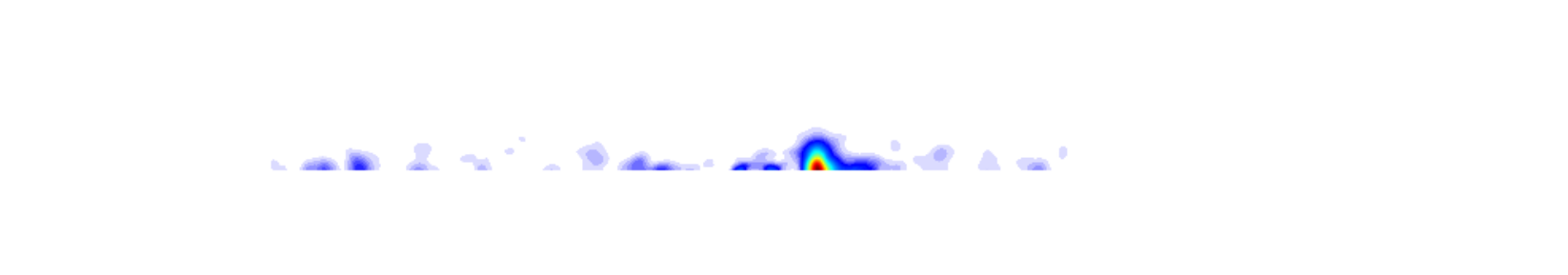}
\put(0,0){\includegraphics[scale=1.0]%
{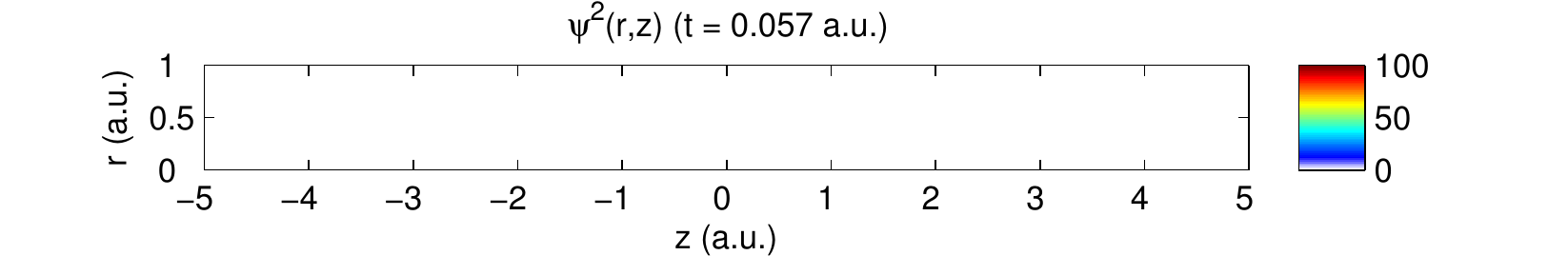}}
\end{overpic} \\
\begin{overpic}[scale=1.0,unit=1\textwidth]
{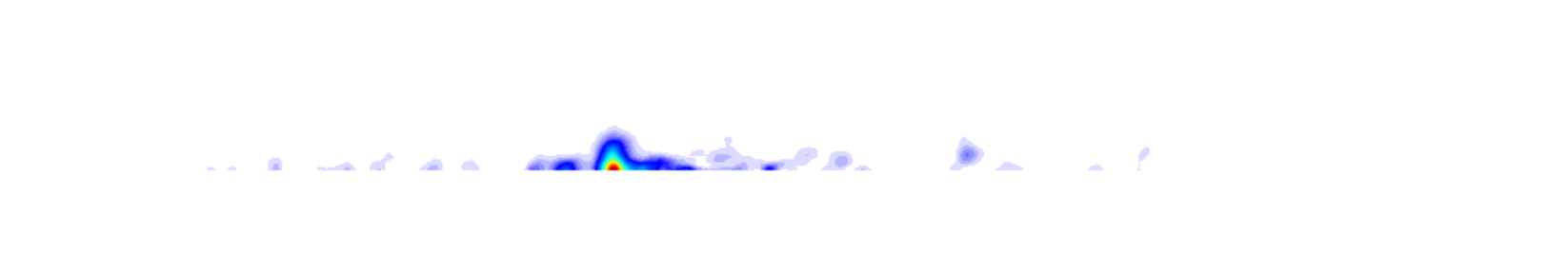}
\put(0,0){\includegraphics[scale=1.0]%
{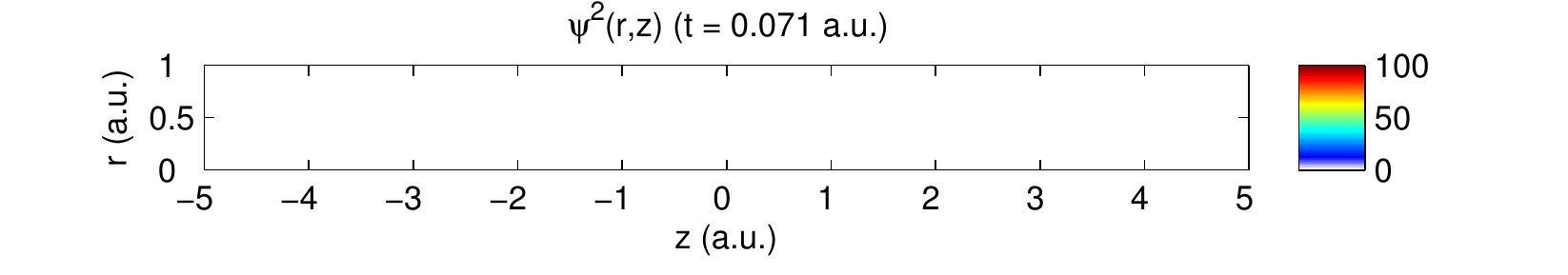}}
\end{overpic}
\caption{Time evolution of the density $|\psi|^{2}$ for an electron in the potential of an hydrogen-like atom (with $Z=10$) in a counterpropagating laser field.}
\label{fig:laser_atom_ev}
\end{figure}

The wave function representing the ground state of an hydrogen-like atom is then evolved in the field of a counterpropagating laser pulse, for which the vector potential is given in Eq. \eqref{eq:pot_laser}. The laser parameters are the same as in Section \ref{sec:3d_gaussian}, except for the maximum electric field which is set to $E = 3.65 \times 10^{5}$ a.u.. The evolution of the wave function is shown in Fig. \ref{fig:laser_atom_ev}. It can be seen in these pictures that the electron moves from left to right, driven by the intense laser field. 


\section{Conclusion}
\label{sec:conclu}

In this work, a numerical method based on a splitting scheme was developed to solve the time-dependent Dirac equation in cylindrical coordinates. The main new idea was the utilization of the Poisson integral solution and Hermite interpolation to solve the radial part of the splitting. This allowed us to circumvent the coordinate singularity problem in $r=0$, while performing all the calculation in coordinate space (no Fourier or Hankel transforms are required): the resulting algorithm is very similar to a non-standard finite difference scheme.

This numerical scheme is an extension of the one presented in \cite{Lorin2011190,FillionGourdeau2012} and thus, shares some of its main properties: 
\begin{enumerate}
\item There is no fermion-doubling problem in the discretization process. Indeed, each step of the splitting is solved in a way that does not modify drastically the continuum dispersion relation, apart from small errors related to the splitting and the radial spatial discretization. A careful analysis of this issue would require a Von Neumann analysis, as in \cite{FillionGourdeau2012} for Cartesian coordinates, where the continuous and discretized Dirac equation are Fourier transformed\footnote{For the radial part, we should use the Hankel transform.} to obtain their respective dispersion relations (similar calculations are found in \cite{FillionGourdeau2012,trefethen:113}). This will be the subject of future investigation.
\item The numerical scheme can be parallelized very efficiently using a domain decomposition method because the algorithm obtained is local in space. Thus, the domain can be separated into subdomains which can be solved independently on different processors: only the information at the subdomain boundaries is exchanged between neighbour processes. This leads to a quasi-linear speedup as the number of processors is increased \cite{FillionGourdeau2012}. This was not considered explicitly in this article, but was verified numerically in some calculation.
\end{enumerate}


The method was tested by looking at the dynamics of wave packets in 2-D. It was shown that the numerical results were in agreement with the analytical result. Moreover, it was tested that the order of convergence of the method in the radial direction is similar to the Cartesian case. Then, a more interesting system was studied which required much more computational resources and where the efficient parallelization of the numerical scheme was required. This allowed us to simulate a 3-D wave packet in a counterpropagating laser field and show the appearance of both negative and positive energy states (interpreted as the creation of antimatter) during the time evolution. 

The numerical method was also combined with the Feit-Fleck spectral scheme and allowed us to evaluate the eigenenergies and eigenfunctions of a 3-D Coulomb-like potential representing a nucleus. Although the performance of the Dirac solver developed in this paper is very good, the calculation of the eigenpairs required a massive amount of computation time. This is mainly due to the slow convergence of the Feit-Fleck method for which the energy resolution scales with the simulation time: a higher resolution requires more calculation times. For this reason, it would be very challenging to make 3-D Quantum Electrodynamics (QED) calculations with the combination of these two methods as the latter necessitates a summation over intermediate states, which involves the computation of all eigenfunctions. However, it may be possible to use a different method to evaluate the initial state which converges faster than the Feit-Fleck method. This is presently under investigation. 

\section*{Acknowledgement}

We would like to thank Huizhong Lu and Szczepan Chelkowski for sharing their work on the numerical solution of the Schr\"odinger equation in cylindrical coordinates.

\appendix

\section{Solution of the Dirac Equation in $z$-coordinates}
\label{app:z_coord}

This Appendix describes precisely how the solution of the Dirac equation in $z$-coordinate is obtained. We are interested first in solving the following equation:
\begin{equation}
 i\partial_{t} \psi(t,r,z) = -i c\alpha_{z} \partial_{z}   \psi(t,r,x)
\end{equation}
with an initial condition given by
\begin{equation}
 \psi(t_{0},r,z) = g(r,z) .
\end{equation}
In the first step, the matrix $\alpha_{z}$ is diagonalized to decouple the spinor components. This is performed by a similarity transformation as
\begin{equation}
 P_{z}^{\dagger}\alpha_{z}P_{z} = \Lambda_{z}
\end{equation}
where $P_{z}$ is a unitary matrix and $\Lambda_{z} = \mathrm{Diag}[1,1,-1,-1] = \beta$ is a diagonal matrix. Starting with $\alpha_{z}$ in the Dirac representation, the explicit expression of the transformation matrix is
\begin{equation}
 P_{z} = \frac{1}{\sqrt{2}} \left(\beta + \alpha_{z} \right).
\end{equation}
The resulting Dirac equation is then 
\begin{equation}
 i\partial_{t} \tilde{\psi} (t,r,z) = -i c\beta \partial_{z}  \tilde{\psi}(t,r,z)
\end{equation}
where we defined $\tilde{\psi} := P^{\dagger} \psi$. It is convenient here, for notational purposes, to split the four-spinor into two bi-spinors as $\tilde{\psi} = (\tilde{\varphi},\tilde{\chi})^{\rm T}$ to get
\begin{eqnarray}
 i\partial_{t} \tilde{\varphi }(t,r,z) &=& -i c \partial_{z}  \tilde{\varphi }(t,r,z) \\ 
 i\partial_{t} \tilde{\chi }(t,r,z) &=& i c \partial_{z}    \tilde{\chi }(t,r,z)
\end{eqnarray}
Therefore, the Dirac equation clearly becomes a set of four uncoupled first-order differential equations in this representation. Their solution is well-known and can be obtained from the method of characteristics. The solution is given by
\begin{eqnarray}
\tilde{\varphi}_{1,2} (t,r,z) &=& \tilde{g}_{1,2}(r,z-c \delta t) , \\
\tilde{\chi}_{1,2}(t,r,z) &=& \tilde{g}_{3,4}(r,z+c \delta t) ,
\end{eqnarray}
along with the conditions $z\pm z_{0} = c \delta t$. The latter is the characteristics along which the partial differential equation becomes an ordinary differential equation. Note that the initial conditions are related to the original representation as $\tilde{g} = P_{z}^{\dagger} g$.

In the last step, the solution is transformed back to the original representation. The final result, after some basic manipulations, is given by
\begin{eqnarray}
 \psi(t,r,z) &=& \frac{1}{2} \left\{ [\mathbb{I}_{4} + \alpha_{z}] g(r,z-c \delta t)   + [\mathbb{I}_{4} - \alpha_{z}] g(r,z+c \delta t)  \right\}.
\end{eqnarray}
This equation is used in the numerical method to evolve the wave function in time in alternate direction iteration.

\section{Solution of the 2D Wave equation}
\label{ann:kirchoff}

The Cauchy problem for the 2-D wave equation in Cartesian coordinates is given by
\begin{eqnarray}
\begin{cases}
 \partial_{t} u(\mathbf{x},t) = c^{2}\nabla^{2}  u(\mathbf{x},t) \\
u(0,\mathbf{x}) := u_{0}(\mathbf{x}), \\
\partial_{t}u(0,t) := v_{0}(\mathbf{x}).
\end{cases} 
\end{eqnarray}
for a general regular $u$. It is well-known that the solution to this 2-D problem is given by the Poisson formula \cite{polyanin2001handbook,evans2010partial}:
\begin{eqnarray}
 u(\mathbf{x},t) &=& \frac{1}{2\pi ct}  \int_{B(\mathbf{x},ct)}d\mathbf{y} \frac{1}{\sqrt{c^{2} t^{2} - |\mathbf{y} - \mathbf{x}|^{2}}} \left[ u_{0}(\mathbf{y}) + tv_{0}(\mathbf{y}) + \nabla u_{0}(\mathbf{y}) \cdot (\mathbf{y} - \mathbf{x}) \right]  
\end{eqnarray}
where $B(\mathbf{x},ct)$ is the ball of radius $ct$ about the position $\mathbf{x}$. Writing this equation in cylindrical coordinates and assuming that the solution describes an axisymmetric system such that the angular dependence can be factorized as $u(r,\theta,t) = e^{i\mu \theta}\tilde{u}(r,t)$, we obtain
\begin{eqnarray}
\label{eq:kirch_cyl}
 \tilde{u}(r,t) &=& \frac{1}{2\pi c t}  \int_{B(r,ct)}RdRd\theta \frac{1}{\sqrt{c^{2} t^{2} - [R^{2} + r^{2} - 2Rr \cos(\theta)]}} \nonumber \\
 && \times \biggl\{ \cos(\mu \theta) \biggl[ \tilde{u}_{0}(R) + t\tilde{v}_{0}(R)   + [R-r\cos(\theta)]\partial_{R}\tilde{u}_{0}(R) \biggr]
  -\sin(\mu \theta) \frac{r}{R}\mu \sin(\theta)\tilde{u}_{0}(R) \biggr\} .
 \end{eqnarray}
This gives an integral representation for the solution in cylindrical coordinates of the wave equation given by:
\begin{eqnarray}
  \partial_{t} \tilde{u}(r,t) = c^{2} \left(\partial_{r}^{2} + \frac{1}{r}\partial_{r} - \frac{\mu^{2}}{r^{2}}  \right) \tilde{u}(r,t).
\end{eqnarray}
This equation is the same as those found in Eq. \eqref{eq:wav_dir1} and therefore, their solutions are given by Eq. \eqref{eq:kirch_cyl}.

\section{Spatial discretization of $\hat{A}$ when $r=0$}
\label{app:A_rzero}

When $r=0$, Eqs. \eqref{eq:proj_sol_dir_cyl1} to \eqref{eq:proj_sol_dir_cyl4} become
\begin{eqnarray}
\label{eq:proj_sol_dir_cyl1r0}
 \psi_{1}^{(1)}(t_{n+1},0,z_{k}) &=& \frac{1}{2\pi a}  \int_{0}^{a}RdR \frac{1}{\sqrt{a^{2} - R^{2}}} \int_{0}^{2\pi}d\theta   \cos(\mu_{1} \theta)\nonumber \\
 && \times \biggl[ g_{1}(R,z_{k}) 
 - a\left[\partial_{R} + \frac{\mu_{2}}{R}\right]g_{4}(R,z_{k}) 
  +R\partial_{R}g_{1}(R,z_{k}) \biggr] ,\\
 \label{eq:proj_sol_dir_cyl2r0}
 \psi_{2}^{(1)}(t_{n+1},0,z_{k}) &=&  \frac{1}{2\pi a}  \int_{0}^{a}RdR \frac{1}{\sqrt{a^{2}  - R^{2}}} \int_{0}^{2\pi}d\theta   \cos(\mu_{2} \theta)\nonumber \\
 && \times  \biggl[ g_{2}(R,z_{k}) 
 -a\left[\partial_{R} - \frac{\mu_{1}}{R}\right]g_{3}(R,z_{k}) 
  + R\partial_{R}g_{2}(R,z_{k}) \biggr] ,\\
 \label{eq:proj_sol_dir_cyl3r0}
 \psi_{3}^{(1)}(t_{n+1},0,z_{k}) &=&  \frac{1}{2\pi a}  \int_{0}^{a}RdR \frac{1}{\sqrt{a^{2}  - R^{2}}} \int_{0}^{2\pi}d\theta   \cos(\mu_{1} \theta)\nonumber \\
 && \times  \biggl[ g_{3}(R,z_{k}) 
 - a\left[\partial_{R} + \frac{\mu_{2}}{R}\right]g_{2}(R,z_{k}) 
  + R\partial_{R}g_{3}(R,z_{k}) \biggr] ,\\
 \label{eq:proj_sol_dir_cyl4r0}
 \psi_{4}^{(1)}(t_{n+1},0,z_{k}) &=&  \frac{1}{2\pi a}  \int_{0}^{a}RdR \frac{1}{\sqrt{a^{2}  - R^{2}}} \int_{0}^{2\pi}d\theta   \cos(\mu_{2} \theta)\nonumber \\
 && \times  \biggl[ g_{4}(R,z_{k}) 
 - a\left[\partial_{R} - \frac{\mu_{1}}{R}\right]g_{1}(R,z_{k}) 
  + R\partial_{R}g_{4}(R,z_{k}) \biggr] .
 \end{eqnarray}
When $\mu_{1},\mu_{2} \neq 0$, then $\psi^{(1)}(t_{n+1},0,z_{k}) = 0$ by virtue of the angular integration, in agreement with the boundary conditions in Eqs. \eqref{eq:exp_form1} to \eqref{eq:exp_form4}. When $j_{z}=\frac{1}{2}$ (the case $j_{z}=-\frac{1}{2}$ is similar), the preceding equations become:
\begin{eqnarray}
\label{eq:psi1_rzero}
\psi_{1,3}^{(1)}(t_{n+1},0,z_{k}) &=& \frac{1}{ a}  \int_{0}^{a}RdR \frac{1}{\sqrt{a^{2} - R^{2}}} \nonumber \\
 && \times \biggl[ g_{1,3}(R,z_{k}) 
 - a\left[\partial_{R} + \frac{1}{R}\right]g_{2,4}(R,z_{k}) 
  +R\partial_{R}g_{1,3}(R,z_{k}) \biggr] ,
 \end{eqnarray}
while $ \psi_{2,4}^{(1)}(t_{n+1},0,z_{k})=0$. Using the same strategy as above, the cubic Hermite interpolation becomes
\begin{eqnarray}
\tilde{g}_{h',1}(r,z_{k})& =&  g_{h',1}(0,k)S_{1}(\xi(r)) + g_{h',1}(1,k)S_{2}(\xi(r))  +   r^{(h')}_{1}m_{h',1}(1,k) S_{4}(\xi(r)), \\
\tilde{g}_{h',2}(r,z_{k})& =&   g_{h',2}(1,k)S_{2}(\xi(r)) + r^{(h')}_{1}m_{h',2}(0,k) S_{3}(\xi(r))+  r^{(h')}_{1}m_{h',2}(1,k) S_{4}(\xi(r)), \\
\tilde{g}_{h',3}(r,z_{k})& =&  g_{h',3}(0,k)S_{1}(\xi(r)) + g_{h',3}(1,k)S_{2}(\xi(r))  +   r^{(h')}_{1}m_{h',3}(1,k) S_{4}(\xi(r)), \\
\tilde{g}_{h',4}(r,z_{k})& =&   g_{h',4}(1,k)S_{2}(\xi(r)) + r^{(h')}_{1}m_{h',4}(0,k) S_{3}(\xi(r))+  r^{(h')}_{1}m_{h',4}(1,k) S_{4}(\xi(r)),
\end{eqnarray}
where $\xi(r) := \frac{r}{r^{(h')}_{1}}$ and where we set $m_{h',1,3}(0,k)=0$ according to the boundary conditions in Eqs. \ref{eq:exp_form1} to \ref{eq:exp_form4}. Also, the remaining derivatives in $r=0$ are approximated by
\begin{eqnarray}
\label{eq:diff1_anti}
m_{h',2,4}(0,k) &\approx & \frac{g_{h'}(1,k)}{2r^{(h')}_{1}}, 
\end{eqnarray}
where we used the fact that the wave function components 1 and 4 are antisymmetric around $r=0$. Thus, with these techniques, the boundary conditions at $r=0$ is fulfilled exactly by the interpolating polynomial.

The next step is the substitution of the interpolant into Eqs. \eqref{eq:proj_sol_dir_cyl1r0} to \eqref{eq:proj_sol_dir_cyl4r0}, resulting in integrals of the form
\begin{eqnarray}
\label{eq:int1_r0}
  \int_{0}^{a}dR \frac{ R^{l}}{\sqrt{a^{2}  - R^{2}}} =  a^{l}\frac{\sqrt{\pi}}{2} \frac{\Gamma\left(\frac{1}{2} + \frac{l}{2} \right)}{\Gamma\left(1+\frac{l}{2} \right)}.
 \end{eqnarray}
It is then possible to evolve the wave function at $r=0$ by using points within the domain, while preserving the boundary conditions. Also, it is clear that we circumvent the coordinate singularity problem as the integration in Eq. \eqref{eq:int1_r0} is well-defined.

\section{Solution of 2-D wave packet}
\label{app:sol_2-D_WP}

In this Appendix, the analytical solution for the time evolution of a 2-D free wave packet with azimuthal symmetry is computed. The solution can be computed in polar coordinates by considering the solution in Cartesian coordinates. We consider an initial wave packet for a massive spin-up electron at rest. The initial wave function in Cartesian coordinates is given by
\begin{equation}
\label{eq:psi_init_2-DWP}
 \psi(t=0, x, y) = \mathcal{N} \begin{bmatrix} 1 \\ 0 \\ 0 \\ 0 \end{bmatrix} e^{- \frac{x^{2} + y^{2}}{4 (\Delta)^{2}}}
\end{equation}
and its Fourier transform by 
\begin{equation}
 \widehat{\psi}(t=0,p_x,p_y) = 4 \pi \Delta^{2} \mathcal{N} 
\begin{bmatrix}
1 \\ 
0 \\ 
0 \\ 
0
\end{bmatrix}
e^{- \Delta^{2} (p_{x}^{2} + p_{y}^{2} ) }.
\end{equation}
The 2-D Dirac equation we want to solve is given by
\begin{equation}
 i\partial_{t} \widehat{\psi}(t,p_{x},p_{y}) = \left[ c\alpha_{x} p_{x} + c\alpha_{y} p_{y}  + \beta m c^{2} \right]\widehat{\psi}(t,p_{x},p_{y}),
\end{equation}
here expressed in Fourier space. The solution to this equation is then simply
\begin{eqnarray}
\widehat{\psi}(t,p_{x},p_{z}) &=& e^{ -ic \alpha_{x} p_{x}t - ic \alpha_{y} p_{y}t -i\beta mc^{2}t } \widehat{\psi}(0,p_{x},p_{y}) \\
&=& \left[ \mathbb{I}_{4} \cos \left( E t \right) -i \frac{c\alpha_{x} p_{x} + c\alpha_{y}p_{y} + \beta mc^{2}}{E} \sin \left(  E t \right) \right] \nonumber \\
&& \times \widehat{\psi}(0,p_{x},p_{y}) ,
\end{eqnarray} 
where $E = \sqrt{p_{x}^{2}c^{2} + p_{y}^{2}c^{2} + m^{2}c^{4}}$. This last equation can be Fourier transformed back to real space. Then, using polar coordinates ($p^{2} = p_{x}^{2} + p_{y}^{2}$) and properties of Bessel functions, we get the solution as
\begin{eqnarray}
\label{eq:WP_an_2-D_1}
 \psi_{1}(t,r,z) &=& 2 \mathcal{N} \Delta^{2} \int_{0}^{\infty} dp p e^{-\Delta^{2}p^{2}} J_{0}(pr)  \nonumber \\ 
&& \times \left [ \cos(Et) - i\frac{mc^{2}}{E}\sin(Et) \right]  \\
\label{eq:WP_an_2-D_2}
 \psi_{2}(t,r,z) &=& 0 \\
\label{eq:WP_an_2-D_3} 
\psi_{3}(t,r,z) &=& 0 \\
\label{eq:WP_an_2-D_4}
 \psi_{4}(t,r,z) &=& 2 \mathcal{N} \Delta^{2} e^{i\theta}\int_{0}^{\infty} dp p J_{1}(pr) \frac{cp}{E} \sin(Et) e^{-\Delta^{2}p^{2}} 
\end{eqnarray}
where $J_{n}(z)$ is the Bessel function of the first kind. Looking at the angular dependence, we see that this solution corresponds to a wave function with $j_{z} = 1/2$.

There is no known analytical form for these integrals in the general case \cite{watson}. However, using high accuracy numerical integration, this result can be used to validate our numerical method and to analyze the operator splitting.

\section*{References}
\bibliographystyle{elsarticle-num}
\bibliography{bibliography}

\end{document}